\documentclass[12pt,preprint]{aastex}

\shorttitle{Sun-to-Earth Propagation of Slow CMEs}

\shortauthors{Liu et al.}

\begin{document}

\title{On Sun-to-Earth Propagation of Coronal Mass Ejections: 2. Slow Events and Comparison with Others}

\author{Ying D. Liu\altaffilmark{1}, Huidong Hu\altaffilmark{1}, Chi Wang\altaffilmark{1}, 
Janet G. Luhmann\altaffilmark{2}, John D. Richardson\altaffilmark{3}, 
Zhongwei Yang\altaffilmark{1}, and Rui Wang\altaffilmark{1}}

\altaffiltext{1}{State Key Laboratory of Space Weather, National Space 
Science Center, Chinese Academy of Sciences, Beijing 100190, China;
liuxying@spaceweather.ac.cn}

\altaffiltext{2}{Space Sciences Laboratory, University of California, Berkeley, 
CA 94720, USA}

\altaffiltext{3}{Kavli Institute for Astrophysics and Space Research, 
Massachusetts Institute of Technology, Cambridge, MA 02139, USA}

\begin{abstract}

As a follow-up study on Sun-to-Earth propagation of fast coronal mass ejections (CMEs), we examine the Sun-to-Earth characteristics of slow CMEs combining heliospheric imaging and in situ observations. Three events of particular interest, the 2010 June 16, 2011 March 25 and 2012 September 25 CMEs, are selected for this study. We compare slow CMEs with fast and intermediate-speed events, and obtain key results complementing the attempt of \citet{liu13} to create a general picture of CME Sun-to-Earth propagation: (1) the Sun-to-Earth propagation of a typical slow CME can be approximately described by two phases, a gradual acceleration out to about 20-30 solar radii, followed by a nearly invariant speed around the average solar wind level; (2) comparison between different types of CMEs indicates that faster CMEs tend to accelerate and decelerate more rapidly and have shorter cessation distances for the acceleration and deceleration; (3) both intermediate-speed and slow CMEs would have a speed comparable to the average solar wind level before reaching 1 AU; (4) slow CMEs have a high potential to interact with other solar wind structures in the Sun-Earth space due to their slow motion, providing critical ingredients to enhance space weather; and (5) the slow CMEs studied here lack strong magnetic fields at the Earth but tend to preserve a flux-rope structure with axis generally perpendicular to the radial direction from the Sun. We also suggest a ``best" strategy for the application of a triangulation concept in determining CME Sun-to-Earth kinematics, which helps to clarify confusions about CME geometry assumptions in the triangulation and to improve CME analysis and observations.   

\end{abstract}


\section{Introduction}

Solar storms, known as coronal mass ejections (CMEs), are massive expulsions of plasma and magnetic flux from the solar atmosphere into interplanetary space. CMEs have been recognized as drivers of major space weather effects. A key question in CME research and space weather forecasting is how CMEs propagate from the Sun all the way to the Earth. Characterizing Sun-to-Earth propagation of CMEs is crucial for at least two aspects: understanding of the physical mechanisms governing CME propagation and interaction with the inner heliosphere; and development of practical capabilities for space weather forecasting. 

Accurate determination of CME Sun-to-Earth kinematics is feasible with the launch of the Solar Terrestrial Relations Observatory \citep[STEREO;][]{kaiser08}. STEREO is comprised of two spacecraft with one preceding the Earth (STEREO A) and the other trailing behind (STEREO B). Each spacecraft carries an identical imaging suite, the Sun Earth Connection Coronal and Heliospheric Investigation \citep[SECCHI;][]{howardra08}, which can image a CME from its birth in the corona all the way to the Earth and beyond. Using a triangulation technique based on the wide-angle imaging observations from STEREO, \citet[][hereinafter referred to as paper 1]{liu13} investigate the Sun-to-Earth propagation of fast CMEs (with speeds larger than 1000 km s$^{-1}$ near the Sun). They find a typical three-phase Sun-to-Earth propagation profile for fast events: an impulsive acceleration, then a rapid deceleration, and finally a nearly constant speed or gradual deceleration. CMEs with different initial speeds are expected to undergo different propagation histories in the heliosphere. A question thus arises regarding how slow CMEs propagate in the Sun-Earth space. 

The first issue, however, concerns the definition of slow CMEs before the question can be addressed. Previous studies suggest that CMEs slower than the ambient solar wind are accelerated whereas events faster than the solar wind are decelerated \citep[e.g.,][]{sheeley99, lindsay99, gopalswamy00}. It is therefore natural to define slow CMEs as those launched with speeds below the average solar wind speed (which is typically around 400 km s$^{-1}$ near the ecliptic). This definition, although appearing trivial, distinguishes slow events from others in terms of the forces acting on them, the Sun-to-Earth propagation profile and heliospheric consequences. First, it implies that a major force speeding up a slow CME would be the solar wind drag \citep[e.g.,][]{cargill04, vrsnak13, zic15}. In contrast, it is purely the Lorentz force that is responsible for the acceleration of fast CMEs. Second, the definition suggests that the maximum speed of a slow CME would be comparable to the ambient solar wind speed, if its acceleration is governed by the solar wind drag. More specifically, a slow CME is expected to be first brought up to about the ambient solar wind speed and then co-move with the solar wind. Yet it is not clear at what distance a slow CME reaches the solar wind speed. Third, it implies heliospheric consequences different from those of fast events. According to this definition a slow CME will not drive a shock and hence will not produce a type II radio burst and energetic particles, unless it propagates into an even slower solar wind environment. Our slow CMEs are similar to the gradual events defined by \citet{sheeley99} as those formed from below streamers with a gradual acceleration, but here we focus on the whole Sun-to-Earth propagation history of CMEs rather than just their acceleration near the Sun. Note that the actual situation of CME interplanetary propagation can be more complicated, involving interactions with the highly structured solar wind and even other CMEs \citep[e.g.,][]{liu12, liu15, lugaz13, isavnin14, rollett14, mishra15}.

The second issue is why slow CMEs matter. First, slow CMEs can also be geo-effective. The CME speed, although a key parameter, contributes to the intensity of geomagnetic storms mainly in the presence of a southward magnetic field. For instance, the most severe geomagnetic storm of the space age (with a minimum $D_{\rm st}$ of $-548$ nT) was produced by a relatively slow CME from 1989 March 10 (with a Sun-to-Earth transit time of 54.8 hr), while the 1972 August 4 CME which has the fastest transit on record (14.6 hr) generated a geomagnetic storm with the minimum $D_{\rm st}$ of only $-125$ nT \citep{cliver90, cliver04}. Second, slow CMEs spend a long time in the Sun-Earth space, which implies a high probability to interact with solar wind structures including other CMEs during transit from the Sun to Earth. Slow CMEs thus provide material, in particular southward magnetic fields and seed particles, that can be processed further in interplanetary space to enhance space weather. Third, the origin of slow CMEs and its implications for their structure and interplanetary propagation are intriguing but not well understood. Slow CMEs often rise as streamer blowouts \citep{sheeley99}. A direct consequence is that these CMEs have to disrupt the streamer as well as the pre-existing heliospheric plasma sheet on their way out. It is not clear how this origin and subsequent interaction with the streamer and heliospheric plasma sheet affect the CME orientation and structure. As will be seen later slow CMEs can also occur as ``stealth" events without low coronal signatures \citep{robbrecht09}, whose structure and evolution are even more difficult to predict. 

In this paper we examine the Sun-to-Earth propagation of slow CMEs. We select three events for this investigation: the 2010 June 16 CME which is a ``stealth" event and exhibited a typical Sun-to-Earth propagation profile of slow CMEs; the 2011 March 25 CME which was significantly deflected towards a head-on impact on the Earth; and the 2012 September 25 CME which was overtaken by a shock near 1 AU enhancing pre-existing southward magnetic fields. Similar to paper 1, this work has a multifold aim but adds a focus on CME structure and geo-effectiveness: (1) to constrain the Sun-to-Earth kinematics of slow CMEs combining heliospheric imaging observations and in situ measurements; (2) to probe the structure of slow CMEs and its implied geo-effectiveness; and (3) to investigate crucial physical processes governing CME propagation and interaction with the heliosphere. We describe observations and methodology in Section 2, present detailed case studies in Section 3, and compare the Sun-to-Earth propagation profile of slow CMEs with those of fast and intermediate-speed events in Section 4. The results are summarized and discussed in Section 5. Our study of slow CMEs provides new insights into CME Sun-to-Earth physics as well as space weather prediction. It also complements the finding of paper 1 on fast CMEs in an attempt to create a general picture of CME Sun-to-Earth propagation. 

\section{Observations and Methodology}

This work requires coordinated heliospheric imaging observations from both STEREO A and B and in situ measurements near the Earth. We use imaging data from the outer coronagraph (COR2) and the heliospheric imagers (HI1 and HI2) of SECCHI aboard STEREO. COR2 has a field of view (FOV) of 0.7$^{\circ}$ - 4$^{\circ}$ around the Sun. HI1 has a 20$^{\circ}$ square FOV centered at 14$^{\circ}$ elongation from the center of the Sun, and HI2 has a 70$^{\circ}$ FOV centered at 53.7$^{\circ}$. HI1 and HI2 can observe CMEs out to the vicinity of the Earth and beyond by using sufficient baffling to eliminate stray light \citep{harrison08, eyles09}. Images from the inner coronagraph (COR1) are also examined but not included here given its small FOV (0.4$^{\circ}$ - 1$^{\circ}$), so the resulting CME kinematics do not include the initiation phase of CMEs. In situ plasma and magnetic field measurements are taken from Wind at L1. Below we describe the methodology for the joint interpretation of the imaging and in situ observations. 

\subsection{Geometric Triangulation of Imaging Observations}

We use a triangulation technique originally proposed by \citet{liu10a} to determine CME Sun-to-Earth kinematics based on the wide-angle imaging observations from STEREO. The technique initially assumes a relatively compact CME structure simultaneously seen by the two spacecraft. It has no free parameters and can give CME kinematics (both propagation direction and radial velocity) as a function of distance from the Sun continuously out to 1 AU. This capability is key to probing CME propagation and interaction with the inner heliosphere. Motivated by this triangulation concept, \citet{lugaz10} and \citet{liu10b} realize that the same idea can be applied by assuming CME geometry as a spherical front attached to the Sun. In this case, what is seen by a spacecraft is the segment tangent to the line of sight. These are essentially the same triangulation concept under different assumptions on CME geometry, because the basic idea, frame, and analysis procedures are all the same as described by \citet{liu10a, liu10b}. Therefore, they are called triangulation with F$\beta$ and HM approximations respectively \citep[see more discussions in][]{liu10b, liu13}. \citet{davies13} developed the same expressions for this triangulation concept using a self-similar model for which the F$\beta$ and HM geometries are limiting cases. The triangulation concept has proved to be a useful tool for determining CME Sun-to-Earth kinematics and connecting imaging observations with in situ signatures \citep[e.g.,][]{liu10a, liu10b, liu11, liu12, liu13, mostl10, lugaz10, harrison12, temmer12, davies13, mishra13}. We will compare the results from the two assumed geometries and see how they work as we move to larger spacecraft longitudinal separations. 

\subsection{In Situ Measurements, Reconstruction and Geo-effectiveness}

We compare the CME kinematics derived from imaging observations with in situ measurements at Wind, focusing on the CME arrival time and speed. In situ measurements can give local plasma and magnetic field parameters along a one-dimensional cut when an interplanetary CME (ICME) encounters a spacecraft. As in paper 1 we use the term ``CMEs" for events observed in images and ``ICMEs" for ejecta identified from in situ measurements. Signatures used to identify ICMEs from solar wind measurements include depressed proton temperatures, smooth and enhanced magnetic fields, and rotation of the field components. We reconstruct the in situ ICME structure using a Grad-Shafranov (GS) technique \citep{hau99, hu02}, which has been validated by well (but not too much) separated multi-spacecraft measurements \citep{liu08a, mostl09}. The GS method relaxes the force-free assumption and can give a cross section as well as flux-rope orientation without prescribing the geometry. The GS reconstruction is sensitive to the chosen ICME boundaries, so it also helps determine ICME intervals. In conjunction with imaging observations it provides a larger spatial perspective of ICMEs than one-dimensional in situ measurements \citep{liu10b}. We also model the $D_{\rm st}$ index using two empirical formulae based on the solar wind data \citep{burton75, om00} and compare the simulation results with $D_{\rm st}$ measurements; the goal is to examine possible geo-effectiveness associated with the ICME structure. 

\section{Case Studies}

We select three events for detailed studies covering some complexity and diversity of the Sun-to-Earth propagation of slow CMEs: the 2010 June 16, 2011 March 25 and 2012 September 25 CMEs. Each of these events has wide-angle imaging coverage from both STEREO A and B and in situ signatures at Wind. The 2010 June 16 CME was relatively isolated, so we can obtain a typical Sun-to-Earth propagation profile of slow CMEs without contamination by other events. The 2011 March 25 and 2012 September 25 CMEs exhibited interactions with a co-rotating interaction region (CIR) and another CME, respectively. The effects of the interactions on the Sun-to-Earth propagation and space weather can be learned from these two cases. Tables~1 and 2 summarize the CME and ICME parameters, which will be described below. 

\subsection{The 2010 June 16 Event}

Figure~1 shows the configuration of the planets and spacecraft in the ecliptic plane on 2010 June 16 when the CME occurred. STEREO A and B were $74.1^{\circ}$ west and $69.5^{\circ}$ east of the Earth at a distance of 0.96 and 1.03 AU from the Sun, respectively. During the 2010 June 16 CME, the Messenger spacecraft was at 0.56 AU from the Sun and $22.3^{\circ}$ east of the Earth. Both Messenger and the Earth were likely impacted by the CME because the CME trajectory was near the Sun-Earth line (also see Table~1). 

Figure~2 displays two synoptic views of the CME (left) from STEREO A and B and the time-elongation maps (right) produced by stacking the running difference images within a slit along the ecliptic \citep[e.g.,][]{sheeley08, davies09, liu10a}. The CME occurred as a streamer blowout with a peak speed of about 390 km s$^{-1}$. There were no visible signatures in the corona (no associated active region, no flare, except a filament that may be partially eruptive), so this appeared to be a ``stealth" event \citep{robbrecht09}. The CME launch time is estimated to be around 08 UT using COR1 observations. The CME was relatively isolated. The track adjacent to the CME front in Figure~2 (right) is a high-density structure trailing the front, which corresponds to the CME core. Another CME occurred about 10 hr later, but it did not seem to affect the propagation of the June 16 event. 

We apply the triangulation technique using elongation angles extracted along the CME track in the time-elongation maps. The resulting CME kinematics in the ecliptic are shown in Figure~3. Both the F$\beta$ and HM approximations give propagation angles east of the Sun-Earth line, consistent with previous estimates \citep{colaninno13, mostl14, shi15}. The propagation angles from the HM approximation are systematically larger and noisier, with the average about 2.4 times that from the F$\beta$ assumption (also see Table~1). The distances and speeds from the two approximations are very similar for the whole elongation range where the triangulation is applied, except that the F$\beta$ approach gives a slight apparent acceleration beyond $\sim$130 solar radii. This late acceleration results from the F$\beta$ assumption of CME geometry and is not physically meaningful \citep[see more discussions in][]{wood09, lugaz09, liu13}. The situation is also seen in the 2011 March 25 and 2012 September 25 cases, and becomes worse when moving to larger longitudinal separations between the two spacecraft (the angle bracketing the Earth). It is worth noting, however, that triangulation with the F$\beta$ approximation yields fairly accurate results when the spacecraft longitudinal separation is smaller than $180^{\circ}$, as illustrated by the present case and other event studies \citep[e.g.,][]{liu10a, liu10b, liu11, mostl10, mishra13}. The latter point was often undervalued previously, and the shortcoming of triangulation with the F$\beta$ geometry tended to be overstated. 

The speed profiles from both the F$\beta$ and HM approximations show a slow acceleration up to 25-30 solar radii and thereafter a roughly constant value at about 390 km s$^{-1}$. This is distinct from the Sun-to-Earth propagation of fast CMEs which typically exhibits three phases: a quick acceleration, then a rapid deceleration, and finally a nearly constant speed or gradual deceleration \citep{liu13}. The terminal speed of 390 km s$^{-1}$ is essentially the average solar wind speed around the ecliptic. It is likely that the CME was primarily accelerated by the forward drag force of the ambient solar wind, so its highest attainable velocity would be the ambient solar wind speed. A similar speed profile was alluded to for another streamer blowout CME by \citet{rollett12} based on a single spacecraft analysis. The predicted speeds at the Earth resulting from the F$\beta$ and HM assumptions are 390 and 360 km s$^{-1}$ respectively (see Table~1), estimated by averaging the data points after 12 UT, June 17. The overlaid X-ray flux does not show any flare associated with the onset of the CME (about 08 UT on June 16). The only flare of June 16 is an A6.9 flare from S25$^{\circ}$W89$^{\circ}$ that peaked around 03:45 UT, which is too west and too early to be associated with this Earth-directed CME. 

Figure~4 shows the in situ counterpart of the 2010 June 16 CME, an isolated ICME without a preceding shock. The ICME leading edge passed Wind at 07:12 UT on June 21 (see Table~2), which is well predicted by both the F$\beta$ and HM triangulations (see Table~1). The total Sun-to-Earth transit time is about 5 days. The speed predictions are also in good agreement with the average speed across the ICME leading boundary (about 394 km s$^{-1}$). The magnetic field within the ICME is not unusually strong. It is unclear whether this is typical for streamer blowout CMEs. There is hardly any southward magnetic field component inside the ICME, suggestive of little geomagnetic storm activity. This result is consistent with both the measured $D_{\rm st}$ and modeled ones using the two empirical formulae \citep{burton75, om00}. Note that we simulate the $D_{\rm st}$ index using the southward magnetic field components in GSM coordinates, although the fields shown in Figure~4 are in RTN coordinates.

The ICME cross section resulting from the GS reconstruction is plotted in Figure~5. Note that the magnetic fields are in a flux-rope frame (with $x$ almost along the spacecraft trajectory and $z$ in the direction of the flux-rope axis). The reconstruction gives a right-handed, complex structure with an axis elevation angle of about $21^{\circ}$ and azimuthal angle of about $285^{\circ}$ in RTN coordinates (see Table~2). These orientation angles are different from, but roughly consistent with, the estimate of \citet{nieves12} derived from a force-free fitting of the measured magnetic fields (note that the orientations at Wind in their Table~2 are in GSE coordinates, not RTN as they stated in the text). This difference is likely owing to the different trailing boundary identified for the ICME, or to the different technique employed for the reconstruction. The angle between our flux-rope orientation and the inclination of the erupted neutral line on the Sun is about $120^{\circ}$, consistent with the interpretation of flux-rope rotation in the corona \citep[and perhaps in interplanetary space as well;][]{vourlidas11}. Similar results on flux-rope rotation have been found before by comparing CME orientation obtained from coronagraph image modeling with GS in situ reconstruction \citep{liu10b}. The Wind spacecraft seems to have crossed an X-like structure separating two magnetic islands, which agrees with the observed depression in the magnetic field strength near the middle of the ICME and considerable rotation in the $B_{\rm R}$ component (see Figure~4). The reconstructed cross section also reveals relatively weak azimuthal components of the flux-rope magnetic field along the spacecraft trajectory. These flux-rope characteristics (e.g., a positive inclination in combination with relatively weak azimuthal field components) may explain the absence of a southward magnetic field and hence nonoccurrence of a geomagnetic storm. 

\subsection{The 2011 March 25 Event}

The 2011 March 25 CME occurred after the two STEREO spacecraft moved to opposite sides of the Sun (see Figure~6). STEREO A and B were $88.5^{\circ}$ and $95.4^{\circ}$ away from the Sun-Earth line, respectively. During the CME Messenger and Mercury were 0.33 AU from the Sun and $51.8^{\circ}$ east of the Earth. The CME trajectory manifests considerable variations. Figure~7 shows the imaging observations from STEREO (left) and the time-elongation maps (right) produced from the running difference intensities within a slit along the ecliptic. This CME is also a streamer blowout event with a peak speed of about 380 km s$^{-1}$ near the Sun. The CME was associated with a filament eruption around 06 UT on March 25 and a flare of unknown magnitude from AR 11176 (S14$^{\circ}$E34$^{\circ}$) that peaked around 06:09 UT. Note the flattening of the CME front in HI1 and finally a concave-outward structure visible in HI2 of STEREO A. This morphology is probably owing to interaction of the CME with the pre-existing heliospheric plasma sheet. Also note another CME that occurred on the opposite side of the Sun almost at the same time. It is unclear if the near-simultaneous occurrence of these two CMEs is a coincidence or falls into the scenario of sympathetic eruptions \citep[e.g.,][]{schrijver11}. If indeed sympathetic it would indicate a long-range magnetic coupling between opposite sides of the Sun, which will be of particular importance for space weather. 

The CME kinematics obtained from the triangulation technique are displayed in Figure~8. The propagation angle from the F$\beta$ approximation starts from about the solar source longitude (E34$^{\circ}$) and then shows considerable changes to finally a direction towards the Earth. The propagation angle from the HM approximation exhibits a similar trend but, again, is systematically larger and noisier. It is unclear whether the change between 12 and 18 UT on March 25 is real or not. Examination of the CME track from STEREO B reveals a kink in the track around that time period (see the right panel of Figure~7). Previous studies of the 2011 March 25 CME obtained a propagation angle of about E30$^{\circ}$ using a graduate cylindrical shell model \citep{colaninno13, shi15}. While this angle is consistent with the initial value from our F$\beta$ triangulation, they did not consider the deflection. The net deflection angle of about 34$^{\circ}$ from the solar source region is comparable to the non-radial ``channelling" (37$^{\circ}$) of the 2014 January 7 CME \citep{mostl15, wang15}. Contrasting with the present case, the 2014 January 7 event moved away from the central meridian of the Sun which led to a side-on collision with the Earth. The 2011 March 25 CME thus provides an important view to space weather as to how an event can be significantly deflected towards a head-on impact with the Earth. The deflection mainly occurred inside 45 solar radii. Examination of the solar disk EUV observations from STEREO B indicates a coronal hole east of the active region. Deflection by the open magnetic fields of the coronal hole \citep[e.g.,][]{gopalswamy09} plus further push by the fast wind from the coronal hole may be responsible for the westward motion of the CME. 

The distances and speeds from the two approximations are very similar out to about 90 solar radii, after which the F$\beta$ triangulation gives an apparent acceleration. Note that STEREO A and B were observing the CME almost from behind the Sun. This late acceleration, again, results from the F$\beta$ assumption of CME geometry but now in combination with the non-optimal observation situation for triangulation. The speed profiles from both the F$\beta$ and HM approximations first increase to about 370 km s$^{-1}$ at a distance of about 20 solar radii, and then turn into another acceleration phase around 35 solar radii with an even lower rate. The second acceleration phase is peculiar but perhaps real, since both the F$\beta$ and HM triangulations give this impression although with different slopes. A possible explanation is as follows: the CME was first accelerated by the forward drag of a typical ambient solar wind possibly plus some Lorentz force, and when the fast wind from the nearby coronal hole caught up with the CME (presumably around 35 solar radii), the Lorentz force (if any) had already ceased and the CME had already reached the typical solar wind speed; the subsequent acceleration thus only came from the fast wind and so was much smaller. This might be the reason that we see two phases of acceleration with different rates. The speed at the Earth predicted by the F$\beta$ and HM triangulations is 450 and 390 km s$^{-1}$ respectively (see Table~1), slightly larger than the observed speed at Wind (377 km s$^{-1}$). The overlaid X-ray flux has a small data gap around the CME onset (06 UT on March 25) but suggests that the associated flare, if any, would be very weak.  

Figure~9 shows the associated ICME with a preceding shock at Wind. It is interesting that even a slow CME like the present one can drive an interplanetary shock. The shock may have formed when the CME was propagating and expanding into an even slower medium (note the upstream speed of only 330 km s$^{-1}$). The shock passed Wind at 15:07 UT on March 29 (see Table~2), which is about 4 hr later and 11 hr earlier than predicted by the F$\beta$ and HM triangulations respectively (see Table~1). The total Sun-to-Earth transit time (about 4.4 days) is shorter than for the 2010 June 16 case, which implies acceleration in interplanetary space (likely due to the fast wind). The ICME interval is fairly large (longer than 2 days), which would not be possible if the CME motion kept the solar source longitude (E34$^{\circ}$). This large interval verifies a head-on collision with the Earth and hence the significant deflection revealed by the triangulation measurements. Note a high-speed stream and a possible CIR following the ICME, which is consistent with the interpretation of further deflection and acceleration by a fast wind as mentioned above. The magnetic field within the ICME, again, is not unusually strong. There is a considerable but fluctuating southward field component starting from March 31. However, no geomagnetic storm occurred in spite of a sudden commencement caused by the shock. This may be partly owing to the low plasma density coincident with the southward fields. According to previous studies, a high density may facilitate ring current intensification by feeding the plasma sheet of the magnetosphere \citep{farrugia06, lavraud06}. The measured $D_{\rm st}$ is generally smaller than the modeled ones.

The GS reconstructed cross section of the ICME is shown in Figure~10, which indicates a crossing close to the flux rope center. The flux rope is right-handed and has an axis elevation angle of about $35^{\circ}$ and azimuthal angle of about $307^{\circ}$ in RTN coordinates (see Table~2). Because of this tilted orientation ($35^{\circ}$) a crossing near the flux rope center by the Earth would not be possible if the CME propagation direction were E34$^{\circ}$. These results, again, confirm a head-on impact with the Earth and thus the deflection from the solar source longitude. The flux-rope orientation is almost perpendicular to the radial direction from the Sun, which is similar to the 2010 June 16 case. The reconstructed results are consistent with a north-to-south rotation of the magnetic field \citep[an NES configuration according to the classification of][]{bothmer98}. The asymmetry in the meridional field component is partly caused by the positive elevation angle of the flux-rope axis. 

\subsection{The 2012 September 25 Event}

The 2012 September 25 CME has been studied by \citet{liu14b} with a focus on interaction with the September 27 event. Here we briefly summarize some of the results relevant to the present work, and then discuss the Sun-to-Earth propagation of the September 25 CME and how its long stay in the Sun-Earth space is relevant to space weather. STEREO A and B were 125.2$^{\circ}$ west and 118.1$^{\circ}$ east of the Earth respectively, so they were observing the CME from behind the Sun. The CME also occurred as a streamer blowout with a peak speed of about 430 km s$^{-1}$. \citet{liu14b} tentatively associate the CME with a C1.1 flare from AR 11575 (N08$^{\circ}$W04$^{\circ}$) that peaked around 09:43 UT on September 25. Near 1 AU the event was overtaken by another CME, which is associated with a long-duration C3.7 flare from AR 11577 (N09$^{\circ}$W31$^{\circ}$) peaking around 23:57 UT on September 27. The interaction between the two CMEs is of particular importance to space weather as it resulted in double strikes on the magnetosphere \citep{liu14b} and elimination of the outer radiation belt \citep{baker13, turner14}.   

Figure~11 displays the CME kinematics. The F$\beta$ triangulation, again, gives propagation angles starting from about the solar source longitude (W04$^{\circ}$), whereas the HM triangulation enlarges the angles and variations. The CME speed profile first increases to about 430 km s$^{-1}$ at a distance between 15 and 20 solar radii, and then shows a slight deceleration before becoming roughly constant. While this speed profile is generally similar to that of the 2010 June 16 CME (see Figure~3), it bears some characteristics of flare-associated CMEs (for instance, a peak speed faster than the ambient solar wind although only slightly, and a deceleration although small). The speed at the Earth predicted by the HM triangulation is 380 km s$^{-1}$ (see Table~1), comparable to but larger than the observed speed at Wind (305 km s$^{-1}$). The overlaid X-ray flux shows a C4.5 flare peaking around 17:53 UT coincident with the CME acceleration, but this is not the associated flare as it is too late and too east (N16$^{\circ}$E72$^{\circ}$). Also shown in Figure~11 are the propagation angle and radial distance of the September 27 CME, which indicates interaction with the September 25 CME near 1 AU. 

The in situ signatures at Wind are shown in Figure~12, which indicate a complex ejecta formed by the interaction between the September 25 and 27 CMEs. A preceding shock passed Wind at 10:16 UT on September 30 (see Table~2), which is about 9.5 hr later than predicted by the HM triangulation (see Table~1). This shock was probably driven by the September 25 CME when it was propagating and expanding into an even slower solar wind environment (note the upstream speed of only 275 km s$^{-1}$). The total Sun-to-Earth transit time of the September 25 CME is about 5 days. Also note another shock inside the ejecta (around 22:19 UT on September 30), which is presumably driven by the September 27 CME. It is this shock that destroyed the outer radiation belt \citep{baker13, turner14}. The second shock was plowing through the preceding CME near 1 AU and compressing its southward magnetic fields. This gives rise to a two-step, intense geomagnetic storm with a global minimum of $-119$ nT \citep[see more discussions in][]{liu14b}. The simulated $D_{\rm st}$ profile using the \citet{om00} formula (minimum $-104$ nT) agrees with actual $D_{\rm st}$ measurements fairly well, except that the global minimum is underestimated. The \citet{burton75} scheme gives a deeper global minimum ($-134$ nT) and a shallower recovery phase than measured. The GS reconstruction gives a left-handed, flux-rope-like structure for the complex ejecta overall and an axis elevation angle of about $13^{\circ}$ and azimuthal angle of about $263^{\circ}$ in RTN coordinates \citep[see Table~2 and][]{liu14b}. It should be stressed that these reconstruction results are likely biased towards the September 27 CME, although the two CMEs could share similar orientations and the same chirality. Because of the low inclination, the geomagnetic storm was caused by the azimuthal components of the flux-rope magnetic fields.   

It is interesting that these two CMEs, whose launch times are separated by more than 60 hr, can still interact in the Sun-Earth space. This is largely owing to the slow motion and hence a prolonged journey of the September 25 CME in the Sun-Earth space. Although only its northern flank was encountered by the Earth \citep{liu14b}, the slow CME provided southward magnetic fields for enhancement in interplanetary space through compression by the later CME/shock, thus increasing geo-effectiveness. The effect of this slow CME indicates the complexity of the Sun-Earth space and the difficulty of accurate space weather prediction, which both point to the extreme necessity to continuously image the whole Sun-Earth space.  

\section{Comparison between Different CMEs}

Here we compare Sun-to-Earth propagation profiles characteristic of slow (with speeds below 400 km s$^{-1}$ as defined here), fast (with speeds above 1000 km s$^{-1}$) and intermediate-speed (with speeds in between) CMEs. We categorize CMEs this way, in a hope to find and compare general pictures or universal features that regulate the Sun-to-Earth propagation of different types of CMEs. However, CME speeds may cover a continuous spectrum and vary depending on the coronal and interplanetary conditions where a CME propagates, so exceptions from the regulations can certainly occur. 

Figure~13 shows the speed-distance profiles of the 2012 March 7 CME (adapted from paper 1), the 2008 December 12 CME \citep[after][]{liu10b} and the 2010 June 16 CME (obtained here), representative of fast, intermediate-speed and slow events respectively. The 2012 March 7 CME with a peak speed of more than 2000 km s$^{-1}$ shows a typical three-phase profile as discovered in paper 1: an impulsive acceleration up to 10-15 solar radii, then a rapid deceleration out to about 50 solar radii, and thereafter a nearly constant speed (or gradual deceleration). The predicted speed at the Earth overestimates the observed speed by about 250 km s$^{-1}$, which may be partly due to the large longitudinal separation between the two STEREO spacecraft ($227^{\circ}$, i.e., both observing the CME from behind the Sun). Note that the observed speed at the Earth is the average solar wind speed in the sheath between the shock and ejecta, which is usually a little smaller than the shock speed at 1 AU. Hence, a better agreement may be achieved if the shock speed (rather than the sheath speed) is used for the comparison. The 2008 December 12 CME with a peak speed of about 700 km s$^{-1}$ has a generally similar profile but different cessation distances for the acceleration and deceleration. Compared with the 2012 March 7 CME, its speed first increases with a lower rate up to about 20 solar radii, then decreases out to 80-90 solar radii, and thereafter becomes roughly constant. This profile resembles the kinematic model assumed by \citet{wood09} to fit their data (with a maximum speed of 689 km s$^{-1}$ and a deceleration cessation distance of 101 solar radii), except that our deceleration is not constant. In contrast, the Sun-to-Earth propagation profile of the 2010 June 16 CME exhibits only two phases as we have seen earlier: an acceleration with an even slower rate up to 25-30 solar radii followed by a nearly invariant speed at about 390 km s$^{-1}$. The predicted speeds at the Earth for the latter two cases are well confirmed by the in situ measurements at 1 AU; the longitudinal separation between the two STEREO spacecraft is $86.3^{\circ}$ for the 2008 December 12 CME and $143.6^{\circ}$ for the 2010 June 16 event.  

As already noted in paper 1, fast CMEs should have a gradual deceleration phase following the rapid deceleration. During the gradual deceleration phase the CME speed is expected to slowly approach the ambient solar wind speed. The subsequent gradual deceleration is a very long process, out to several tens of AU from the Sun, as the energy is slowly dissipated into the ambient medium (also see Section 4.1 of paper 1). Because of this long process we can assume a nearly constant speed between the cessation distance of the rapid deceleration and 1 AU. A good example is the 2006 December 13 CME, which has a projected speed of 1770 km s$^{-1}$ near the Sun as measured by LASCO aboard SOHO, a shock speed of about 1030 km s$^{-1}$ at the Earth measured in situ by ACE, and a shock speed of about 870 km s$^{-1}$ measured by Ulysses at 2.73 AU \citep{liu08b}. The shock speed at the Earth is comparable to its counterpart at 2.73 AU but significantly smaller than the speed near the Sun. Therefore, the primary deceleration must have occurred inside 1 AU, and further out the event moved with a roughly constant speed. \citet{liu08b} employ a kinematic model, which assumes a constant deceleration followed by an invariant speed, to fit the frequency drift of the long-duration type II burst associated with the event and the 1 AU shock parameters simultaneously. Their best fit gives a deceleration cessation distance of about 78 solar radii. Note that their model does not include an acceleration phase, so their deceleration profile is essentially an average over the acceleration and rapid deceleration stages. These results, obtained with a different approach by \citet{liu08b}, are generally consistent with our triangulation measurements presented here and in paper 1.

The cessation distance of CME rapid deceleration (about 50 and 85 solar radii for the 2012 March 7 and 2008 December 12 CMEs respectively) is much shorter than the average cessation distance of 0.76 AU (163 solar radii) inferred indirectly by \citet{gopalswamy01}. This was already noticed for fast CMEs in paper 1, and now we add evidence by including more events. \citet{winslow15} assume the average transit speed between the Sun and Mercury as the ICME speed at Mercury (due to lack of solar wind plasma measurements there), and by comparing the assumed speed with the average ICME speed at 1 AU they argue that the CME deceleration continues beyond the orbit of Mercury (on average 0.38 AU or 82 solar radii) and may stop around 1 AU. Note that neither \citet{gopalswamy01} nor \citet{winslow15} have or examine wide-angle heliospheric imaging observations, so they cannot follow a CME continuously in the Sun-Earth space and thus cannot determine the deceleration cessation distance precisely. In particular, the assumption of the transit speed as the CME speed at Mercury made by \citet{winslow15} may considerably overestimate the actual CME speed at Mercury, because the major deceleration takes place relatively close to the Sun. For example, the 2012 March 7 CME has an average transit speed of about 1250 km s$^{-1}$ from the Sun to 80 solar radii, which is larger than the speed at 80 solar radii by (at least) 250 km s$^{-1}$ (not even considering the non-optimal observation geometry for triangulation). Therefore, their assumption essentially ``moves" some of the deceleration from within Mercury's orbit to beyond, or in another word ``forces" the major deceleration, which actually stops before Mercury's orbit, to last longer. 

\section{Conclusions and Discussion}

We have investigated the Sun-to-Earth propagation of slow CMEs, combining heliospheric imaging observations from STEREO and in situ measurements at Wind. The 2010 June 16, 2011 March 25 and 2012 September 25 CMEs, which show particularly interesting Sun-to-Earth characteristics, are carefully studied. Our methodology includes a triangulation technique that enables determination of CME kinematics through the Sun-Earth space, GS reconstruction that gives ICME structure, and $D_{\rm st}$ examination that indicates geo-effectiveness associated with the ICME structure. The results complement paper 1 in the attempt to create a general picture of CME Sun-to-Earth propagation, shed light on CME structure and geo-effectiveness, and also have important implications for future CME observations and data analysis. 

\subsection{CME Sun-to-Earth Characteristics}

Key findings are obtained on how a typical slow CME propagates through the Sun-Earth space. Our case studies also cover complications involving interactions with a CIR and another CME, respectively. Below we formulate CME Sun-to-Earth propagation by including both slow and fast events, and discuss the physical mechanisms controlling the Sun-to-Earth propagation and implications for space weather. Note that further work is needed on how CMEs of all speeds propagate through, and interact with, the inner heliosphere between the Sun and Earth. We have barely scratched the surface of this important problem in the field. Anyhow this work will be of interest because it addresses a few of what are the great majority of CMEs.  

1. The Sun-to-Earth propagation of a typical slow CME can be approximately described by two phases: a gradual acceleration out to about 20-30 solar radii, followed by a nearly invariant speed around the average solar wind level. This behavior is largely found in the 2010 June 16 and 2012 September 25 cases, but the 2011 March 25 event also gives such an indication. For the 2011 March 25 case, we observe a second, slow acceleration phase, which may be caused by the impetus from a fast solar wind stream catching up with the CME. As expected, slow CMEs co-move with the solar wind after they have accelerated up to the typical solar wind speed. This two-phase propagation history complements the three-phase profile of fast CMEs found in paper 1, and now we have a general picture regulating the Sun-to-Earth propagation of slow and fast events respectively. As discussed earlier, fast CMEs are purely accelerated by the Lorentz force, whereas slow CMEs are predominantly ``dragged up" in speed by the viscosity force of the ambient solar wind (i.e., carried outward by the solar wind) although the Lorentz force might also contribute. 

2. The comparison of Sun-to-Earth propagation profiles between different types of CMEs indicates that faster CMEs tend to accelerate and decelerate more rapidly and have shorter cessation distances for the acceleration and deceleration (see Figure~13). The more rapid acceleration and shorter acceleration distance/time can be understood in the context of more free energy which is released impulsively from the active region. This scenario is found in CMEs associated with long-duration flares. CME deceleration is usually explained by a viscous drag force (as solar gravity can be neglected), which is proportional to a drag coefficient, the CME cross section, the ambient solar wind density and the square of the velocity difference between the CME and the solar wind \citep[e.g.,][]{cargill04, vrsnak13, zic15}. While the standard form of the drag force is consistent with the observed non-uniform deceleration, it is unclear if it can account for the more rapid deceleration and shorter deceleration distance of faster CMEs. A major problem is that it does not consider the CME-driven shock sweeping up material in the solar wind and producing energetic particles. First, the shock has a cross section much larger than that of the CME. The shock can significantly accrete, heat and accelerate the upstream solar wind. Second, a significant portion of the energy can go into energetic particles through shock acceleration as we envision in paper 1. This point of view seems to be supported by the finding of \citet{mewaldt08} that the total energy content of energetic particles can be 10\% or more of the CME kinetic energy. Therefore, the majority of the momentum and energy loss for fast CMEs is expected to occur via the shock, not the CME directly. More specifically, a fast CME may not be simply ``dragged down" by the viscosity force, but would rather be decelerated by the momentum and energy loss to the ambient medium through the shock.  

3. The comparison between different types of CMEs also suggests that both intermediate-speed and slow CMEs would have a speed comparable to the average solar wind level before reaching 1 AU. This result is important for space weather forecasting: the impact speed of those CMEs at the Earth can be predicted fairly accurately by simply using the average solar wind speed; in particular, for slow CMEs the arrival time at the Earth can also be well predicted with knowledge of the speed and arrival time at $\sim$25 solar radii (as the speed would be roughly constant thereafter). Note that no modeling effort is needed in these predictions. Clearly these findings help facilitate space weather forecasting.

4. Slow CMEs have a high potential to interact with other solar wind structures in the Sun-Earth space due to their slow motion, providing critical ingredients to enhance space weather. The CMEs studied here have a Sun-to-Earth transit time typically around 5 days, during which quite a few other CMEs and CIRs can develop and pass through the Sun-Earth distance. The 2011 March 25 CME was likely deflected by a coronal hole and the subsequent fast stream/CIR from E34$^{\circ}$ towards a head-on impact on the Earth. Although no geomagnetic storm occurred, it reveals how a significant longitudinal deflection can arise to have a head-on collision with the Earth. The 2012 September 25 event was overtaken by the shock of another CME near 1 AU although their launch times are more than 60 hr apart. This led to enhancement of the pre-existing southward magnetic fields and a consequent intense geomagnetic storm. These essentially conform to the ``perfect storm" scenario proposed by \citet{liu14a}, i.e., a combination of circumstances results in aggravation of the situation. This particular view of slow CMEs revealed here further reinforces the conjecture that ``perfect storms" can be frequent enough for us to worry about \citep{liu15}.  

5. The slow CMEs studied here lack strong magnetic fields at the Earth but tend to preserve a flux-rope structure with axis generally perpendicular to the radial direction from the Sun. These characteristics are likely reflections of the origin of the CMEs, which are all streamer blowout events. Although a rigorous proof is not feasible here, we speculate that the accumulated energy beneath the streamer is not big enough to achieve an unusually strong magnetic field inside the ejecta (and a high speed as well), and any filaments involved may simply be tracers of the shearing magnetic field topology rather than causes. In this sense, slow CMEs may be more of an evolutionary transient of a gradually evolving corona. The low speed and lack of strong magnetic fields seem to suggest that streamer blowout CMEs are generally not effective by themselves in producing intense geomagnetic storms, even though they can be geo-effective in combination with other structures (as discussed above). However, geo-effectiveness is also determined by where an observer crosses a structure, so it can be very geo-effective in some places and weakly so in others. Despite the fact that the flux rope can rotate in the corona and interplanetary space, the perpendicularity of the flux-rope axis to the radial direction from the Sun tends to be preserved. These results also illustrate a useful approach that combines flux-rope reconstruction and $D_{\rm st}$ examination, as in \citet{liu15}, to understanding how the plasma and magnetic field characteristics of CMEs are connected with geomagnetic storm intensity and variability. 

\subsection{Implications for CME Triangulation}

The unique merit of our triangulation concept lies in the fact that it can determine CME kinematics as a function of distance throughout the whole Sun-Earth space, not just near the Sun. This feature is an improvement over other techniques. In a separate paper we will evaluate this triangulation concept by comparing the two assumed CME geometries (i.e., F$\beta$ and HM) in a statistical sense. However, there are confusions or misunderstandings about the advantages and disadvantages of the two geometries assumed in the triangulation, which needs an immediate clarification. The following insights obtained from this work will clarify the confusions and help improve CME observational strategies and analysis.  

The propagation angles given by the HM triangulation usually show large variations and tend to significantly deviate from the solar source longitude at distances near the Sun. These characteristics can hardly be true. It is necessary to reiterate this point here, because CME propagation direction is an important aspect in CME kinematics as well as space weather prediction. As already noticed in paper 1, the F$\beta$ triangulation yields more reliable propagation angles at distances near the Sun than the HM triangulation, while further out the opposite seems true. The HM geometry, which assumes a spherical front attached to the Sun, may considerably overestimate the CME size near the Sun (implying an angular width of $180^{\circ}$); far way from the Sun the F$\beta$ geometry is too simple to represent the real CME structure (although we can argue that the feature being tracked is the dominant structure in the CME). These effects are probably the reason for the foregoing issue. We tentatively gave a cutoff distance of 50-100 solar radii in paper 1, but this distance may certainly depend on the CME size and the longitudinal separation between the two STEREO spacecraft.  

We, again, confirm that the F$\beta$ triangulation generally shows a good performance when the two spacecraft observe Earth-directed CMEs before the Sun, despite its appearance of oversimplification. Unphysical acceleration arises at large elongations when the two spacecraft are behind the Sun, and becomes worse with increasing longitudinal separation between the spacecraft (the angle bracketing the Earth). There could also be a slight, unphysical late acceleration when the two spacecraft are before the Sun, but the results from the F$\beta$ triangulation are generally satisfactory in this case. The HM triangulation can suppress the unphysical late acceleration as we see here and in paper 1. However, the HM geometry brings about complications such as permitting multiple solutions, which may be difficult to handle using the HM triangulation alone \citep[also see discussions in][]{liu10b}. 

It also comes into our attention that where to take elongation measurements along the CME track in a time-elongation map may affect the CME arrival time prediction. We usually choose the black/white boundary, which is in general the easiest because the contrast there is the sharpest. For some cases complications may arise (e.g., track bifurcation), so we follow the front edge of the CME track instead. Picking the front edge or the black/white boundary can shift the predicted arrival time. One may expect that the difference is about half the duration of the CME sheath (not necessarily half the sheath duration at 1 AU), since the time-elongation map is constructed from running-difference images. However, extrapolation using a linear or second-order polynomial fit of the distances might change the difference when the CME cannot be tracked to 1 AU by the triangulation technique. To select the front edge or the black/white boundary should be guided by whether a least ambiguous tracking of the CME feature is ensured.

Based on the above discussions we suggest a ``best" strategy for using the triangulation technique: (1) apply the F$\beta$ triangulation first given its simpleness and use the results to pick the right solution from the HM triangulation; (2) adopt the propagation angles from the F$\beta$ triangulation for distances near the Sun and the angles from the HM triangulation for large distances (a tentatively suggested cutoff distance is somewhere between 50-100 solar radii but may vary); (3) adopt the distances and speeds from the HM triangulation when unphysical late acceleration is observed in the F$\beta$ triangulation, but a comparison between the two on the predicted arrival time is always good; (4) implications of where elongations are measured along the CME track may be worth taking into account for the predicted arrival time. Application of the triangulation technique, which is expected to be a routine possibility in the future when dedicated spacecraft are available at L4 and L5 \citep{liu10b}, should be guided by this strategy and its subsequent possibly improved version.  

\acknowledgments The research was supported by the Recruitment Program of Global Experts of China, NSFC under grant 41374173 and the Specialized Research Fund for State Key Laboratories of China. We acknowledge the use of data from STEREO, Wind and GOES and the $D_{\rm st}$ index from WDC in Kyoto.

\clearpage

\begin{deluxetable}{lcccccc}
\tabletypesize{\footnotesize}
\tablecaption{Estimated CME Parameters from Imaging Observations}
\tablewidth{0pt}
\tablehead{
\colhead{Event} & \colhead{AR\tablenotemark{a}} & \colhead{$t_0$\tablenotemark{b}} & 
\colhead{$v_0$\tablenotemark{b}} & \colhead{$\bar{\beta}$\tablenotemark{c}(F$\beta$/HM)} & 
\colhead{$t_{\rm p}$\tablenotemark{d}(F$\beta$/HM)} & 
\colhead{$v_{\rm p}$\tablenotemark{d}(F$\beta$/HM)} \\
 & & (UT) & (km s$^{-1}$) & ($^{\circ}$) & (UT) & (km s$^{-1}$) 
}
\startdata
2010 Jun 16   &-                           &08       &390 &$-6.8$/$-16.4$  &Jun 21 06:05/Jun 21 09:03  &390/360 \\
2011 Mar 25   &1176 (S14E34)    &06       &380 &$-5.3$/$-7.5$    &Mar 29 11:18/Mar 30 02:25  &450/390 \\
2012 Sept 25 &1575 (N08W04)   &09:40  &430 &7.7/11.8 &-/Sep 30 00:43\tablenotemark{e}  &-/380\tablenotemark{e} \\
\enddata
\tablenotetext{a}{Associated active region on the Sun.}
\tablenotetext{b}{Estimated launch time and maximum speed near the Sun.}
\tablenotetext{c}{Average propagation angle relative to the Sun-Earth line, positive if west and negative if east.}
\tablenotetext{d}{Predicted arrival time and speed at Wind derived from F$\beta$ and HM triangulations, respectively. When CME tracking is not possible for the whole Sun-Earth distance, we estimate the arrival time using a linear fit of the distances and the arrival speed by averaging the speeds after they become roughly constant. We have used $r\cos\beta$ and $v\cos\beta$ for the HM geometry in calculating the arrival time and speed, where $r$ is the distance from the Sun, $v$ the speed, and $\beta$ the propagation angle with respect to the Sun-Earth line.}
\tablenotetext{e}{Only predictions from the HM triangulation are shown here due to the non-optimal observation geometry for the F$\beta$ triangulation.}
\end{deluxetable}

\clearpage

\begin{deluxetable}{lccccccc}
\tabletypesize{\small}
\tablecaption{Estimated ICME Parameters from In Situ Measurements}
\tablewidth{0pt}
\tablehead{
\colhead{Event} & \colhead{$v_{\rm o}$\tablenotemark{a}} & \colhead{Shock\tablenotemark{b}} 
& \colhead{Start\tablenotemark{b}} & \colhead{End\tablenotemark{b}} & \colhead{$\theta$\tablenotemark{c}} 
& \colhead{$\phi$\tablenotemark{c}} & \colhead{Chirality} \\
& (km s$^{-1}$) & (UT) & (UT)  & (UT) & ($^{\circ}$) & ($^{\circ}$) & 
}
\startdata
2010 Jun 16  &394  &-                      &Jun 21 07:12  &Jun 22 10:05 &21 &285  &R \\
2011 Mar 25  &377  &Mar 29 15:07  &Mar 29 22:05  &Apr 1 05:46   &35 &307  &R \\
2012 Sep 25 &305  &Sept 30 10:16 &Sept 30 12:50 &Oct 2 00:14\tablenotemark{d}   
&13\tablenotemark{d} &263\tablenotemark{d}  &L\tablenotemark{d} \\
\enddata
\tablenotetext{a}{Observed speed at Wind, which is the average speed in the sheath between the shock and ejecta or across the ICME leading boundary if there is no shock.}
\tablenotetext{b}{Shock arrival time and ICME boundaries at Wind.}
\tablenotetext{c}{Flux-rope axis elevation and azimuthal angles in RTN coordinates, respectively.}
\tablenotetext{d}{The rear boundary and GS reconstruction results correspond to the whole complex ejecta formed by the merging of the 2012 September 25 and 27 CMEs.}
\end{deluxetable}

\clearpage

\begin{figure}
\epsscale{0.8} \plotone{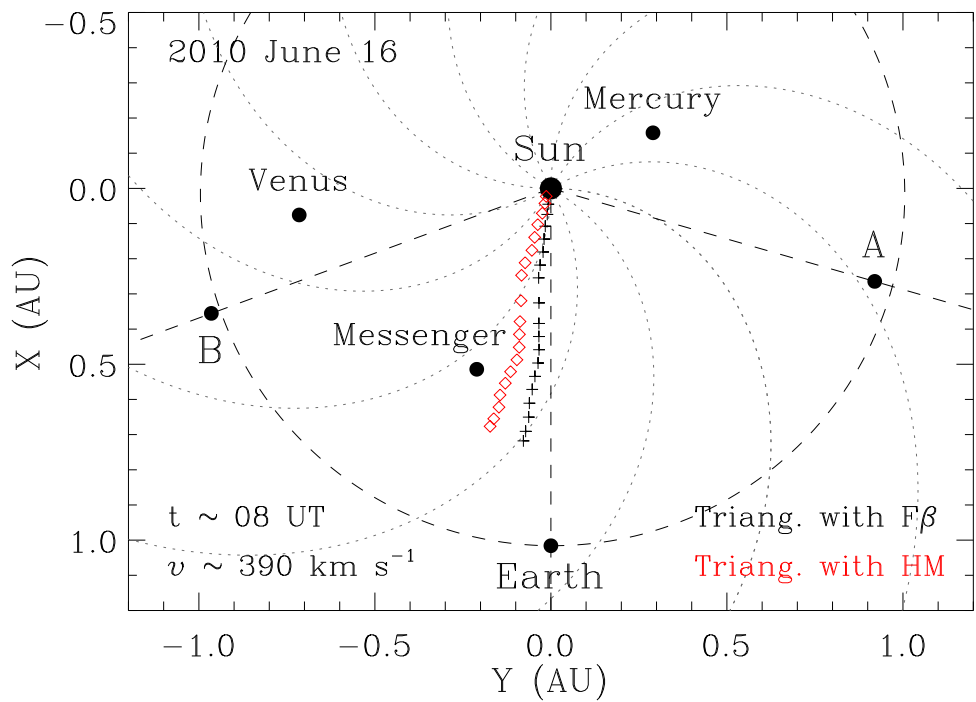} 
\caption{Positions of the spacecraft and planets in the ecliptic plane on 2010 June 16. The CME trajectory is obtained by triangulation with the F$\beta$ approximation (black crosses) and triangulation with the HM approximation (red diamonds), respectively. The estimated CME launch time on the Sun and maximum speed are also given. The dashed circle indicates the orbit of the Earth, and the dotted lines show the spiral interplanetary magnetic fields created with a solar wind speed of 450 km s$^{-1}$.}
\end{figure}

\clearpage

\begin{figure}
\centerline{\includegraphics[width=17pc]{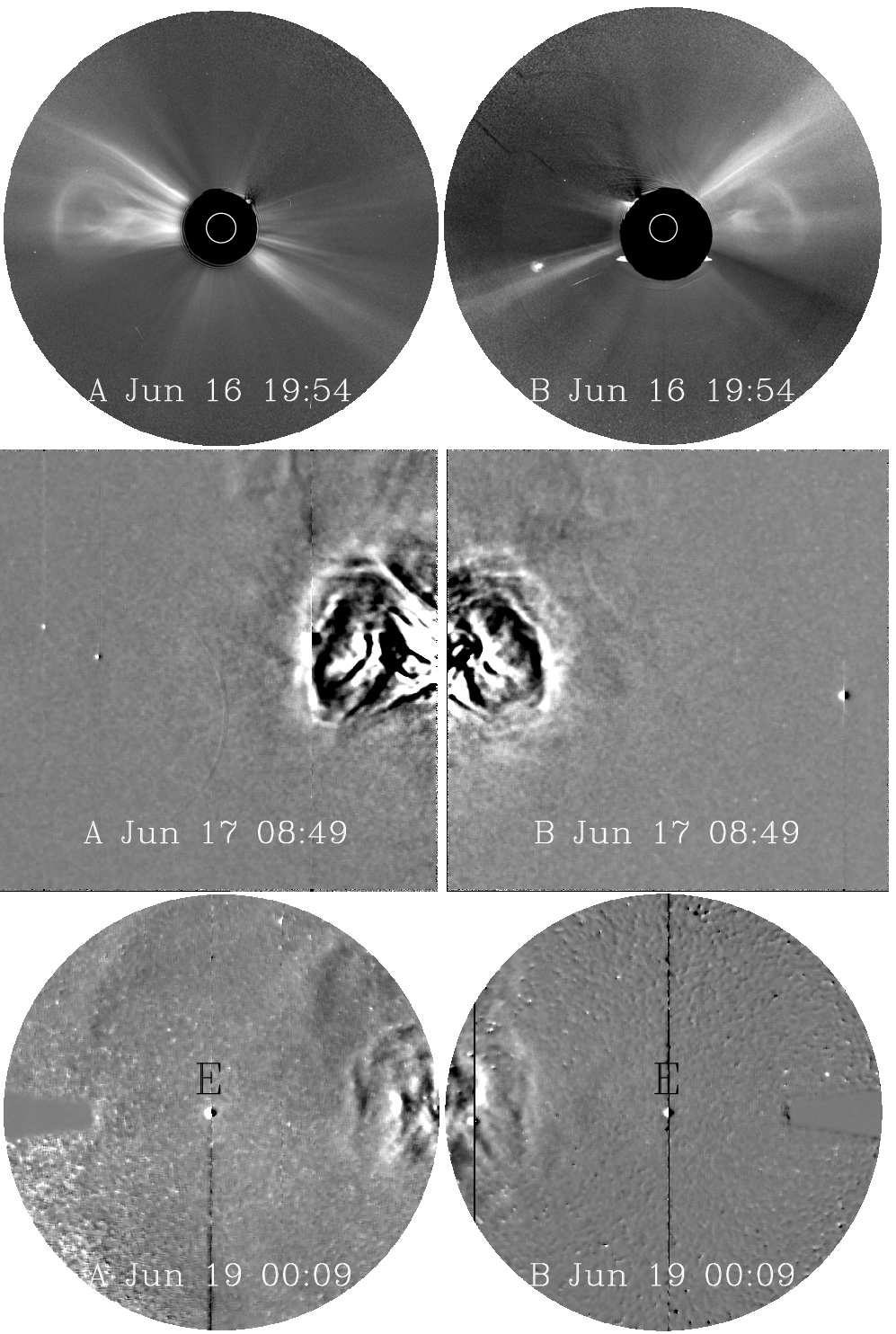}\hspace{0.6pc}\includegraphics[width=20pc]{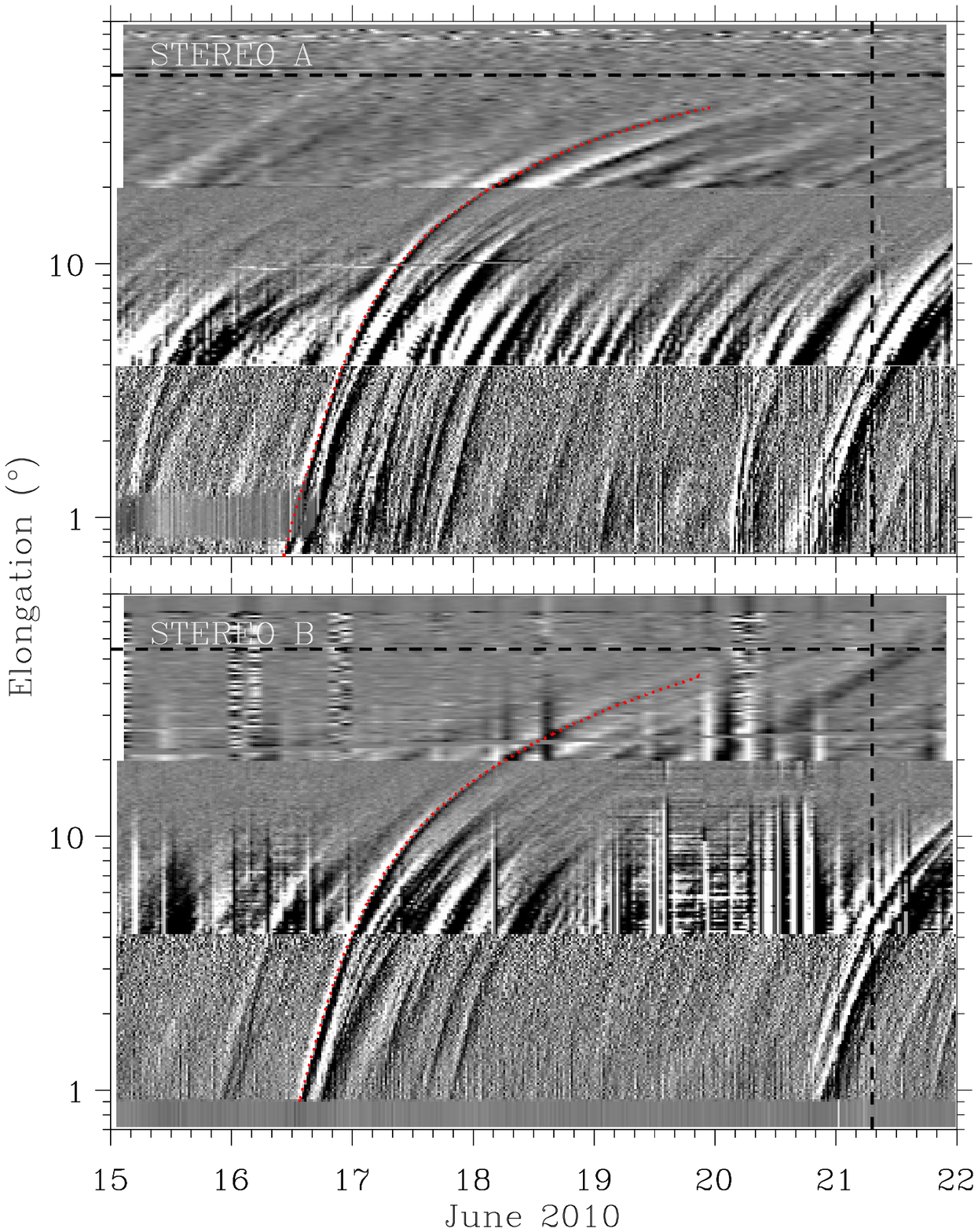}}
\caption{Left: Evolution of the 2010 June 16 CME viewed from STEREO A and B near simultaneously. From top to bottom, the panels show the images of COR2 and running difference images of HI1 and HI2, respectively. The position of the Earth (E) is labeled in the HI2 images. Right: Time-elongation maps constructed from running-difference images along the ecliptic. The red dotted curve indicates the CME track, along which the elongation angles are extracted. The vertical dashed line marks the observed arrival time of the ICME leading edge at the Earth, and the horizontal dashed line denotes the elongation angle of the Earth.}
\end{figure}

\clearpage

\begin{figure}
\epsscale{0.75} \plotone{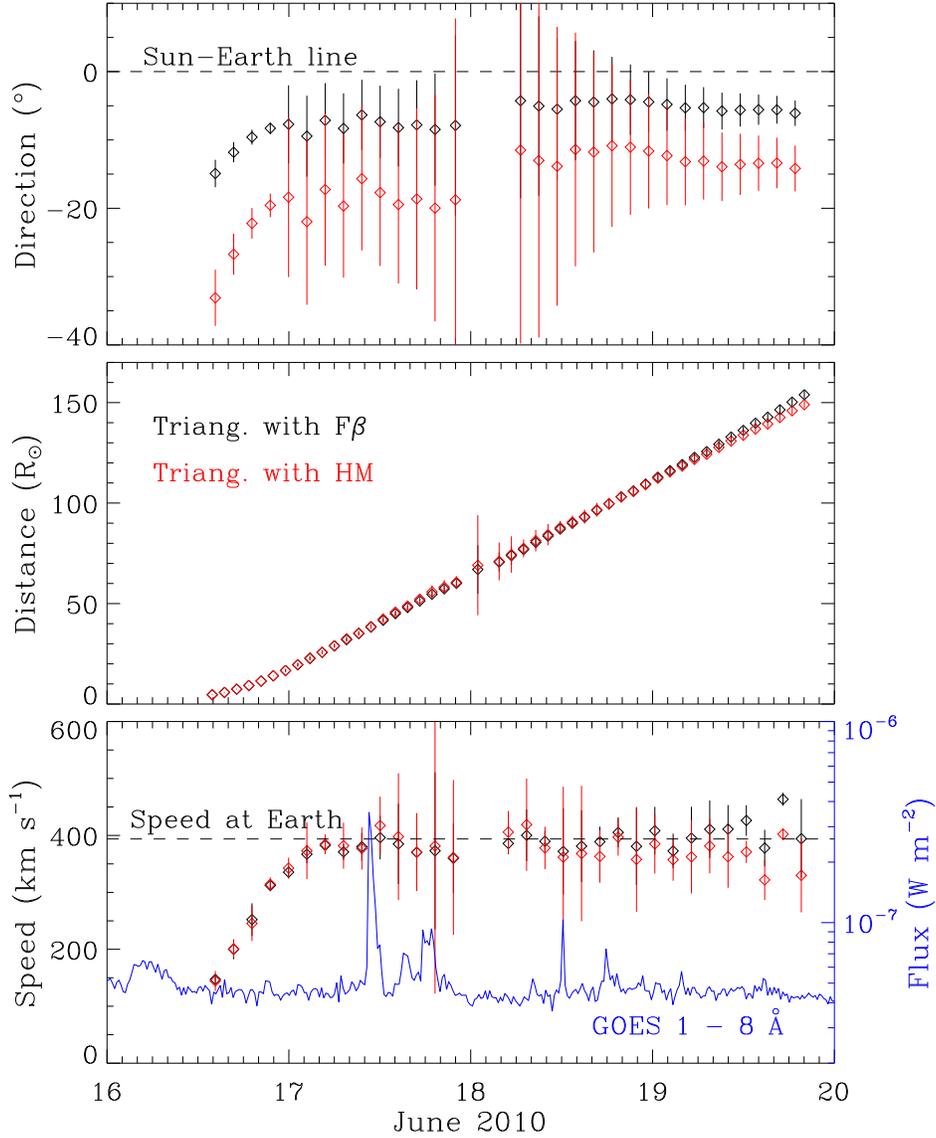} 
\caption{Kinematics of the leading edge of the 2010 June 16 CME derived from triangulation with F$\beta$ (black) and HM (red) approximations. The dashed line in the top panel indicates the longitude of the Earth, while the dashed line in the bottom panel marks the average solar wind speed across the ICME leading edge observed in situ near the Earth. The CME speeds are computed from adjacent distances using a numerical differentiation technique and are then binned to reduce scattering. Overlaid on the speeds is the GOES X-ray flux (scaled by the blue axis). Note a data gap due to singularities in the calculation scheme caused by the spacecraft longitudinal separation angle \citep{liu11}.}
\end{figure}

\clearpage

\begin{figure}
\epsscale{0.75} \plotone{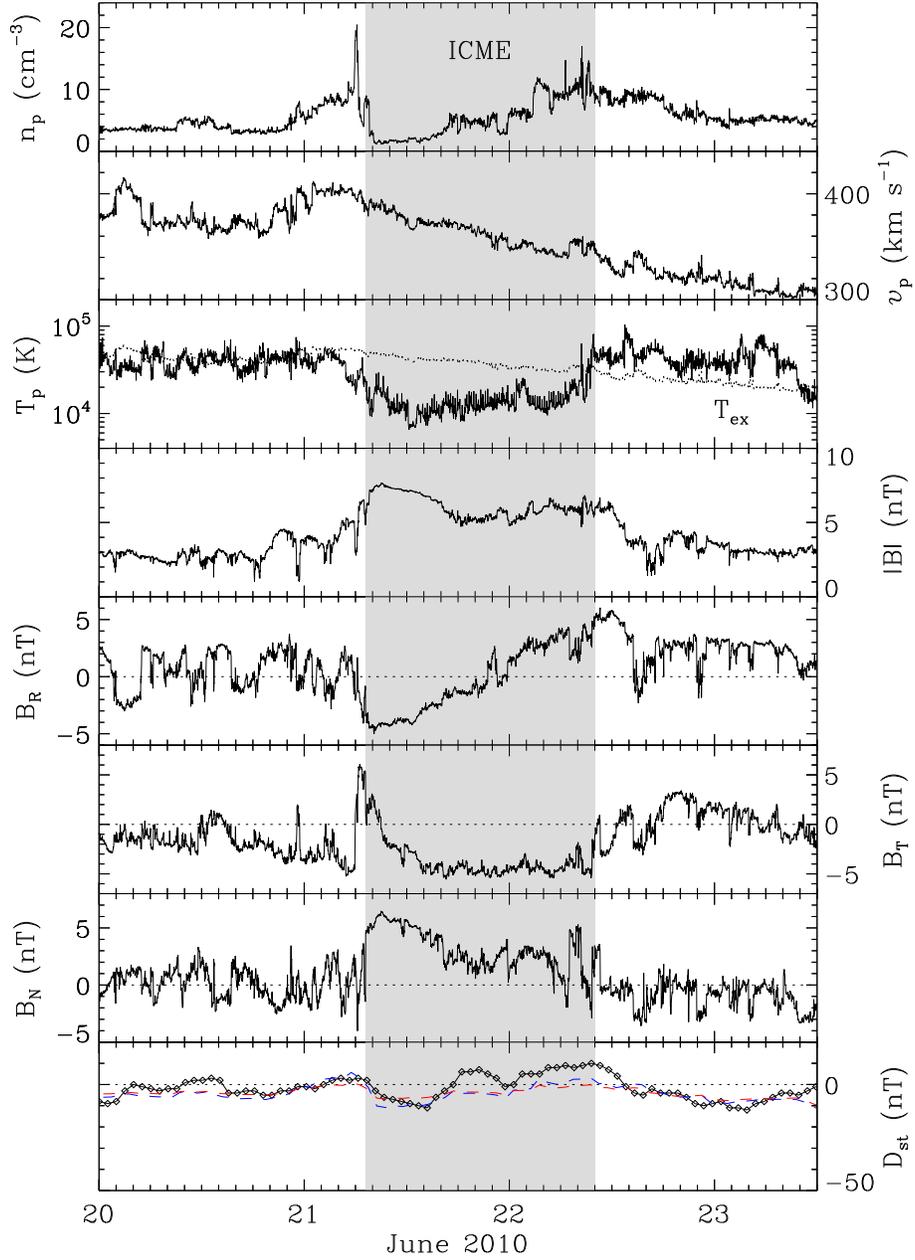} 
\caption{Solar wind measurements at Wind and associated $D_{\rm st}$ index for the 2010 June 16 CME. From top to bottom, the panels show the proton density, bulk speed, proton temperature, magnetic field strength and components, and $D_{\rm st}$ index, respectively. The shaded region indicates the ICME interval. The dotted curve in the third panel denotes the expected proton temperature calculated from the observed speed \citep{lopez87}. The red and blue curves in the bottom panel represent $D_{\rm st}$ values estimated using the formulae of \citet{om00} and \citet{burton75}, respectively.}
\end{figure}

\clearpage

\begin{figure}
\epsscale{0.9} \plotone{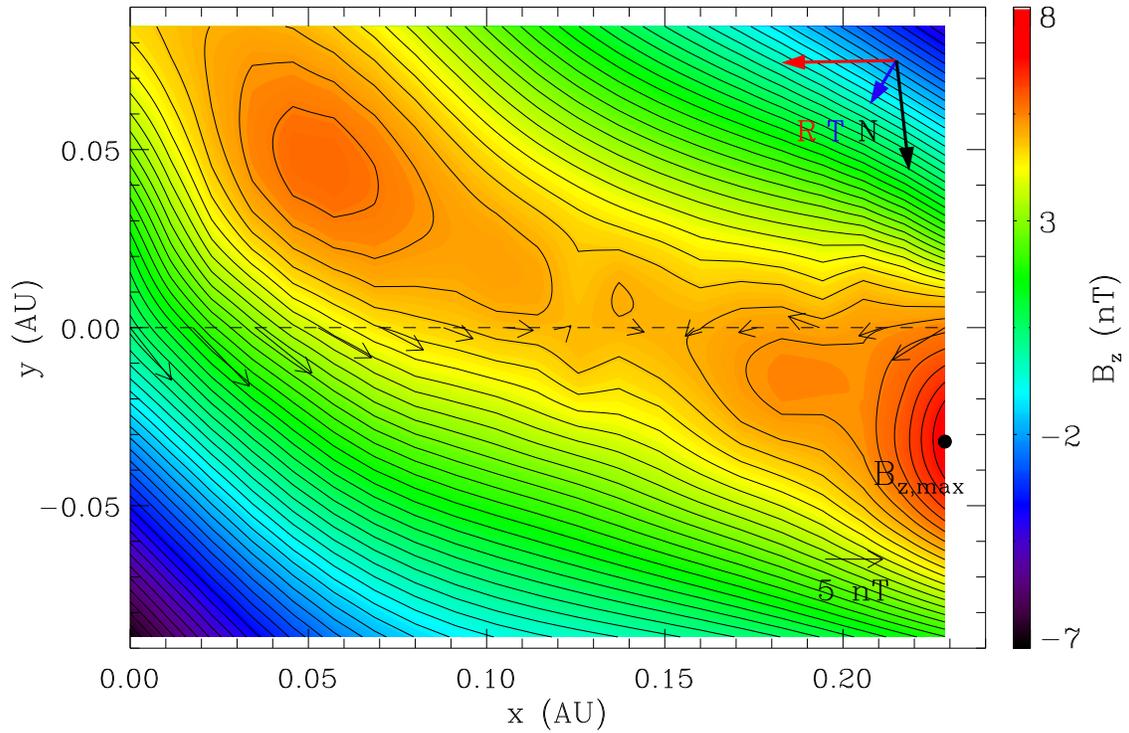} 
\caption{Reconstructed cross section of the 2010 June 21 ICME. The color shading shows the value of the axial magnetic field, and the black contours are the distribution of the vector potential. The location of the maximum axial field is indicated by the black dot. The dashed line marks the trajectory of the Wind spacecraft. The thin black arrows denote the direction and magnitude of the observed magnetic fields projected onto the cross section, and the thick colored arrows represent the projected RTN directions.}
\end{figure}

\clearpage

\begin{figure}
\epsscale{0.8} \plotone{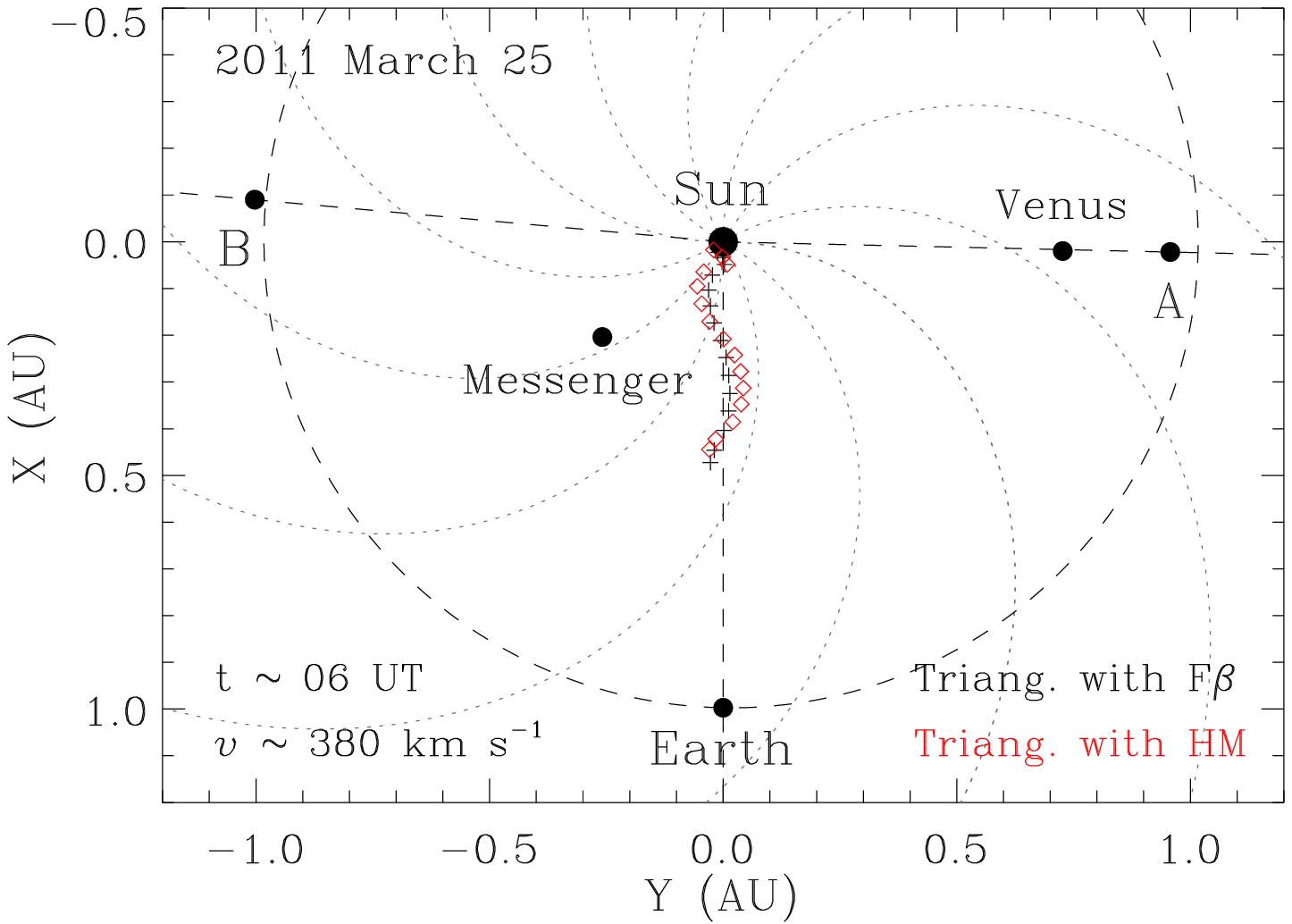} 
\caption{Positions of the spacecraft and planets in the ecliptic plane on 2011 March 25. Similar to Figure~1. Note that the two STEREO spacecraft were separated by more than 180$^{\circ}$ in longitude (the angle bracketing the Earth).}
\end{figure}

\clearpage

\begin{figure}
\centerline{\includegraphics[width=17pc]{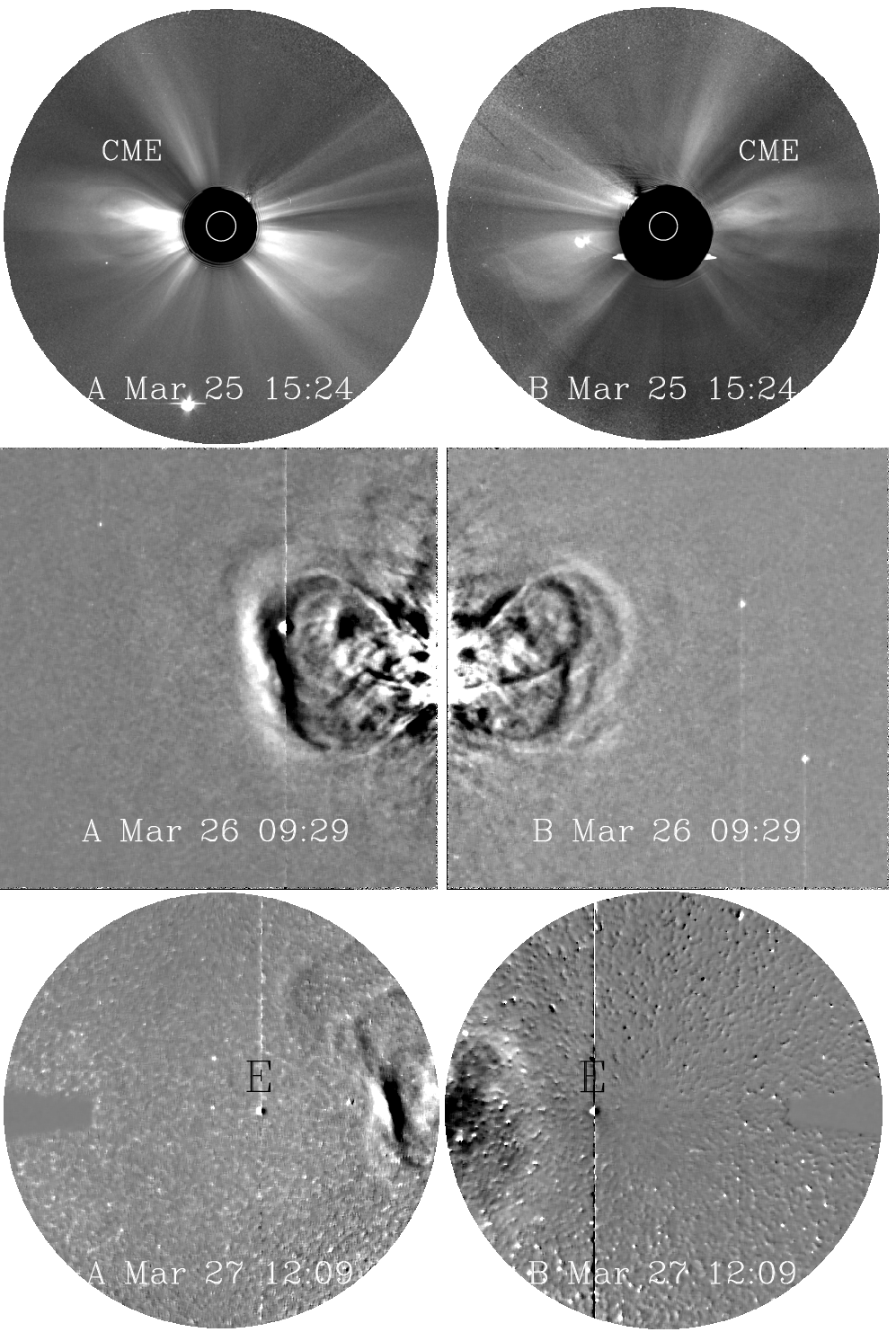}\hspace{0.6pc}\includegraphics[width=20pc]{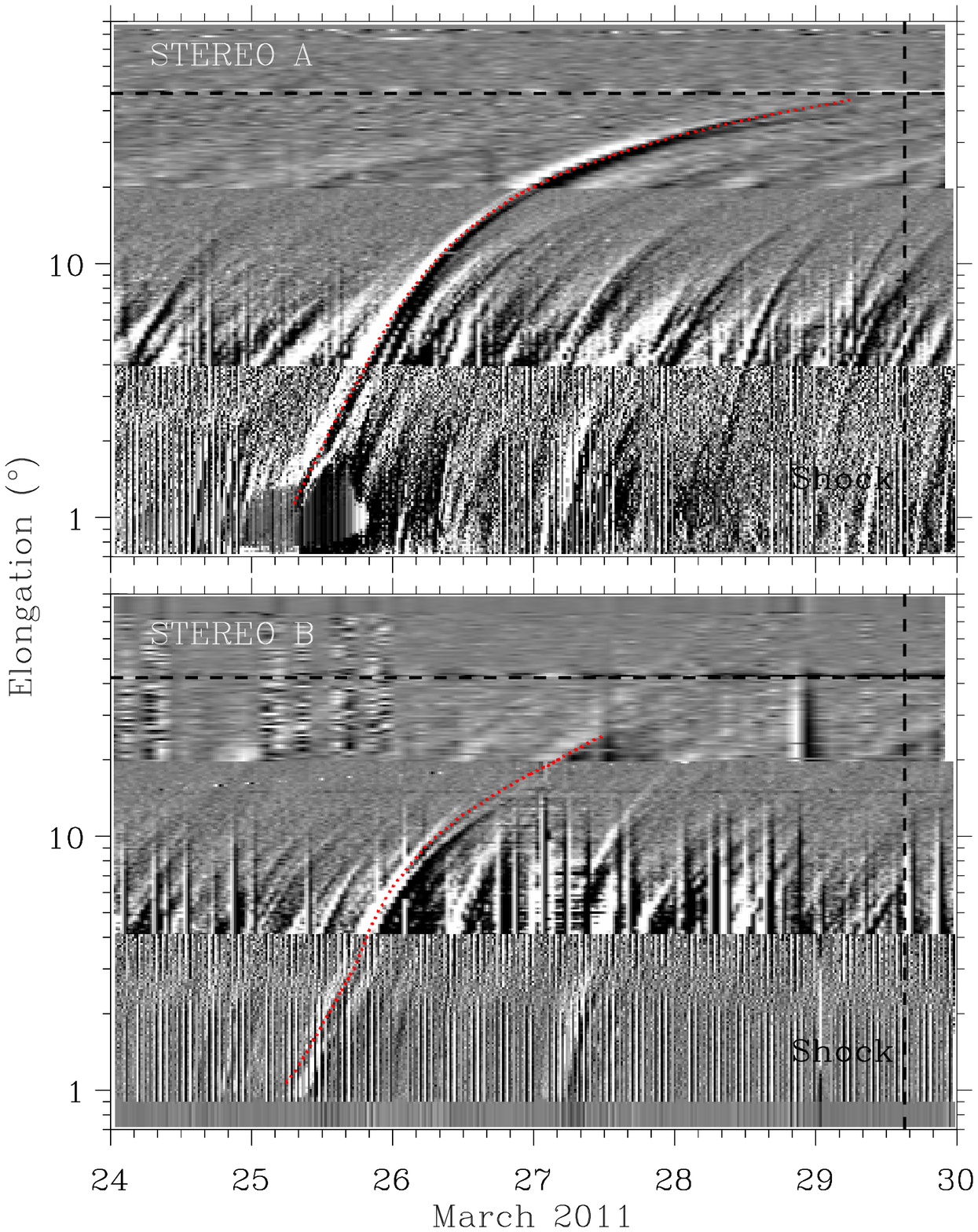}}
\caption{Left: Evolution of the 2011 March 25 CME viewed from STEREO A and B near simultaneously. The CME of interest is marked in COR2, but note another event occurring almost at the same time on the opposite side of the Sun. Right: Time-elongation maps constructed from running-difference images along the ecliptic. Similar to Figure~2. Note that the track behind the red curve corresponds to the CME core. The vertical dashed line indicates the observed arrival time of the CME-driven shock at the Earth.}
\end{figure}

\clearpage

\begin{figure}
\epsscale{0.75} \plotone{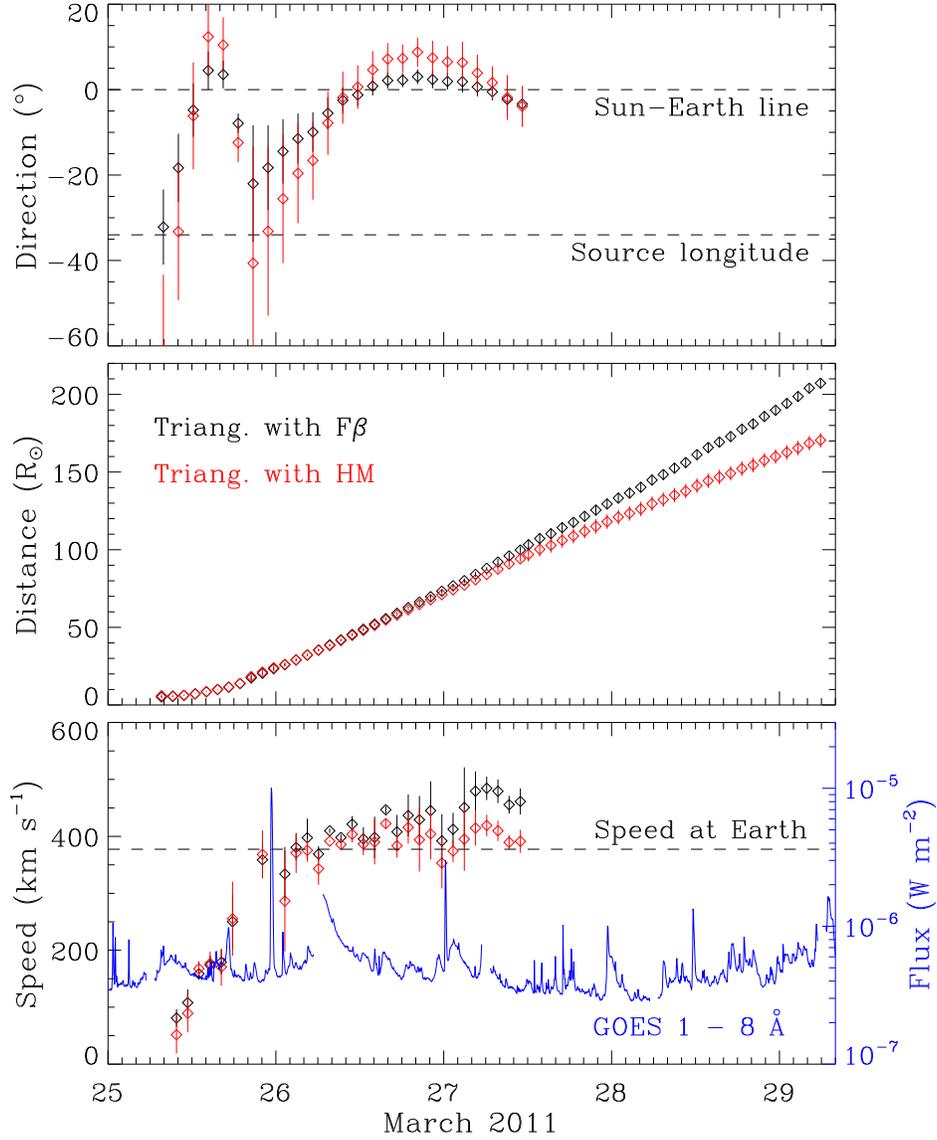} 
\caption{Kinematics of the leading edge of the 2011 March 25 CME derived from triangulation with F$\beta$ (black) and HM (red) approximations. Similar to Figure~3. The longitude of the CME source location on the Sun is also indicated in the top panel, in addition to the Sun-Earth line. The dashed line in the bottom panel marks the average solar wind speed in the sheath region behind the shock observed in situ near the Earth. Elongation measurements after 12:01 UT on March 27 are available only from STEREO A, so the distances thereafter are calculated from STEREO A observations assuming a propagation angle of $-4^{\circ}$ and $-5^{\circ}$ for the F$\beta$ and HM geometries, respectively.}
\end{figure}

\clearpage

\begin{figure}
\epsscale{0.75} \plotone{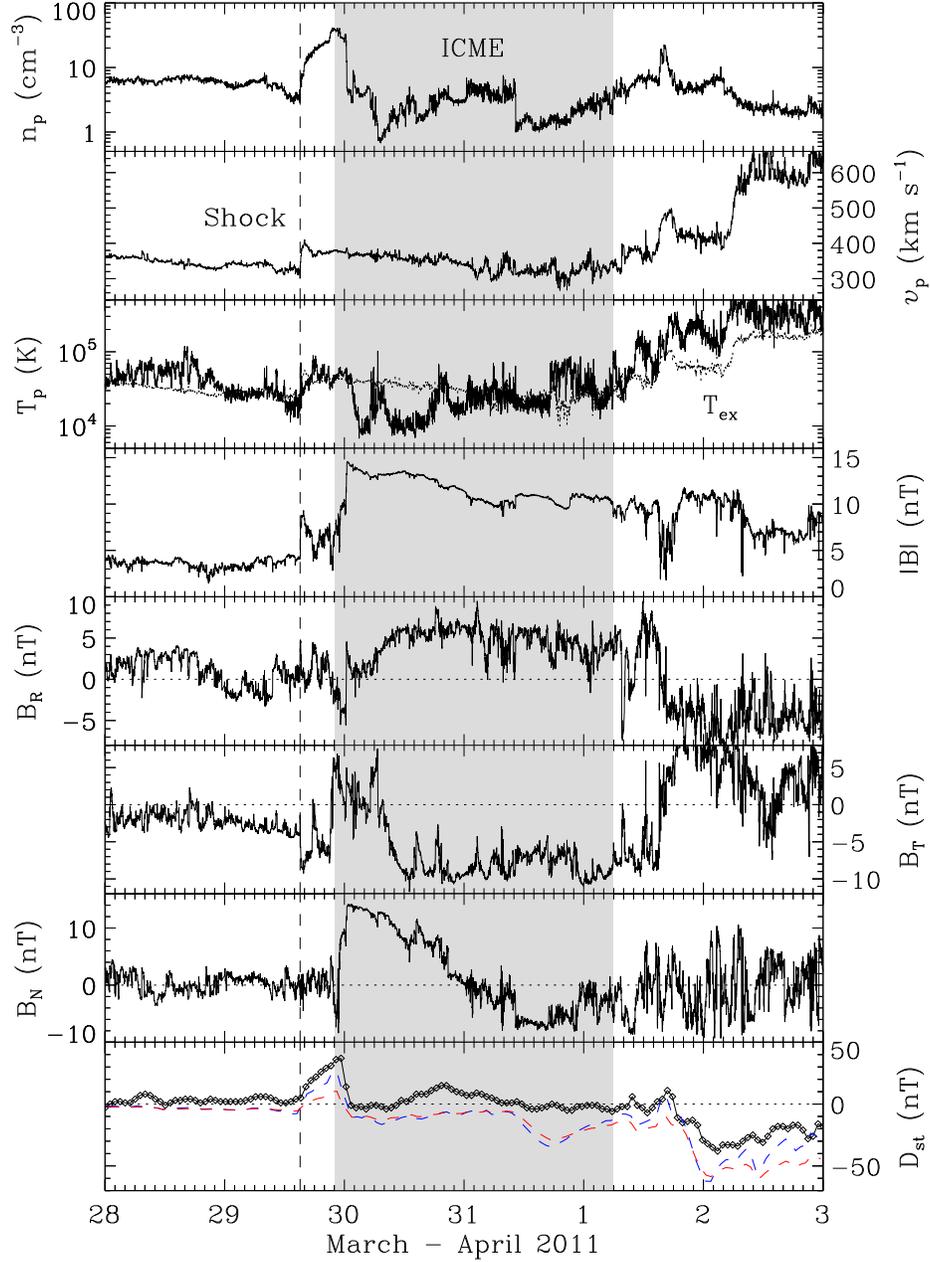} 
\caption{Solar wind measurements at Wind and associated $D_{\rm st}$ index for the 2011 March 25 CME. Similar to Figure~4. The vertical dashed line indicates the CME-driven shock. Note a long ICME interval despite a relatively small CME.}
\end{figure}

\clearpage

\begin{figure}
\epsscale{0.9} \plotone{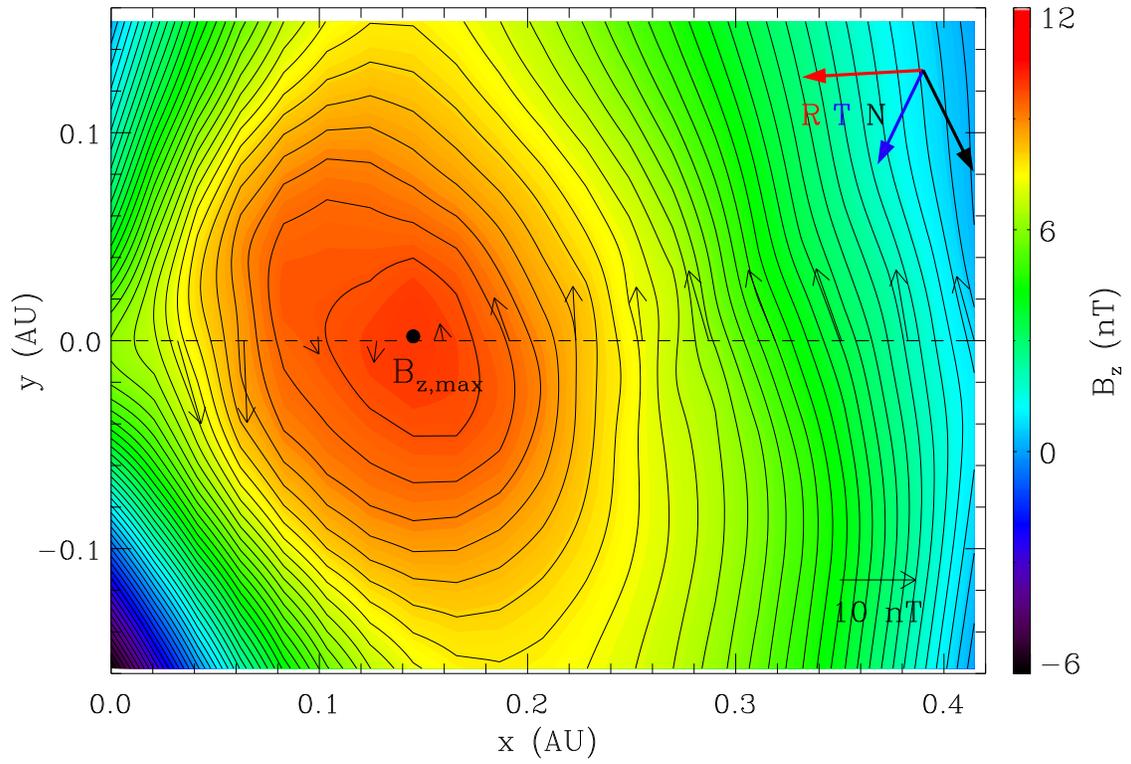} 
\caption{Reconstructed cross section of the 2011 March 29 ICME. Similar to Figure~5.}
\end{figure}

\clearpage

\begin{figure}
\epsscale{0.75} \plotone{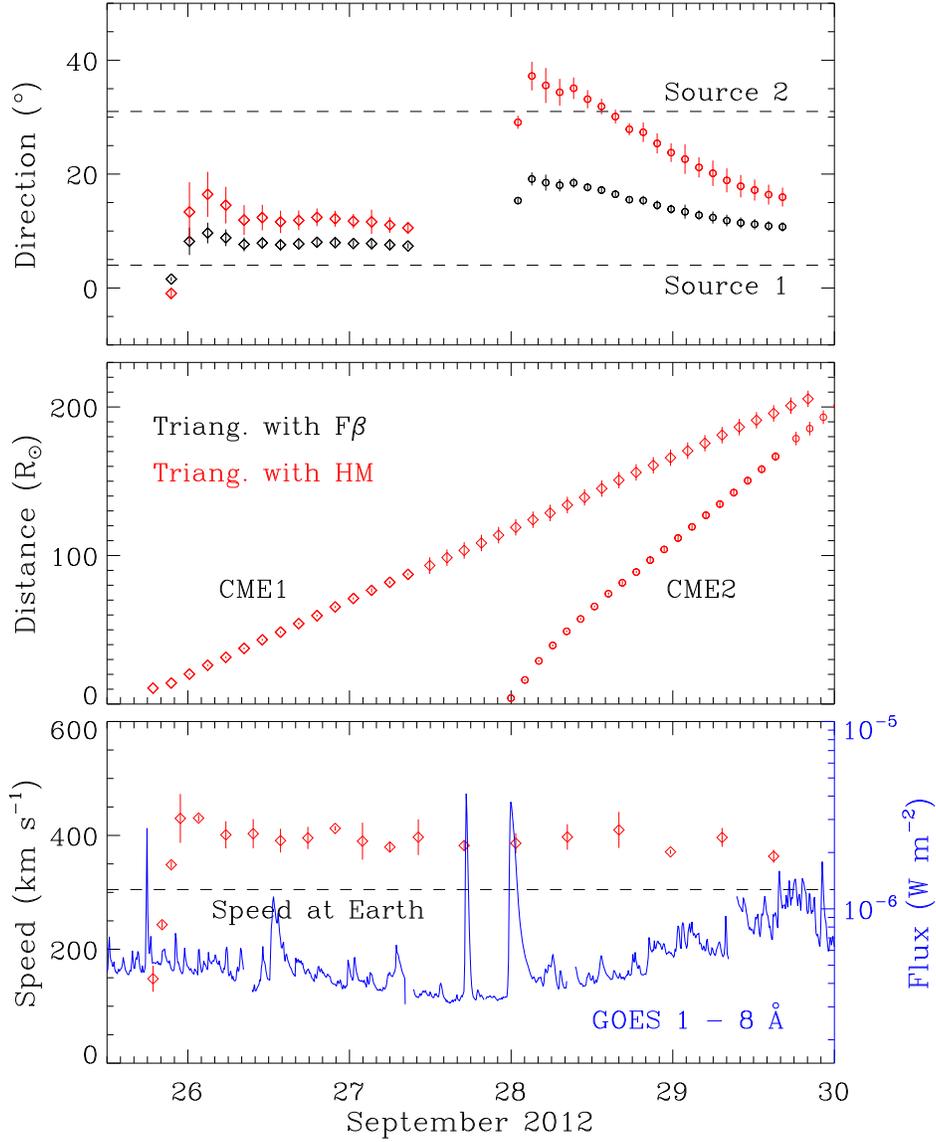} 
\caption{Kinematics of the leading edge of the 2012 September 25 CME \citep[adapted from][]{liu14b}. Similar to Figure~3. The acceleration phase is not fully recovered partly because of a gap in STEREO B observations. Also shown are the propagation angle and radial distance of another event (CME2), which occurred on September 27 and overtook the September 25 CME (CME1) near the Earth \citep{liu14b}. Distances and speeds only from the HM triangulation are plotted due to the non-optimal observation geometry of the two spacecraft for the F$\beta$ triangulation. The dashed line in the bottom panel marks the average solar wind speed in the sheath region behind the first shock observed near the Earth (see Figure~12).}
\end{figure}

\clearpage

\begin{figure}
\epsscale{0.75} \plotone{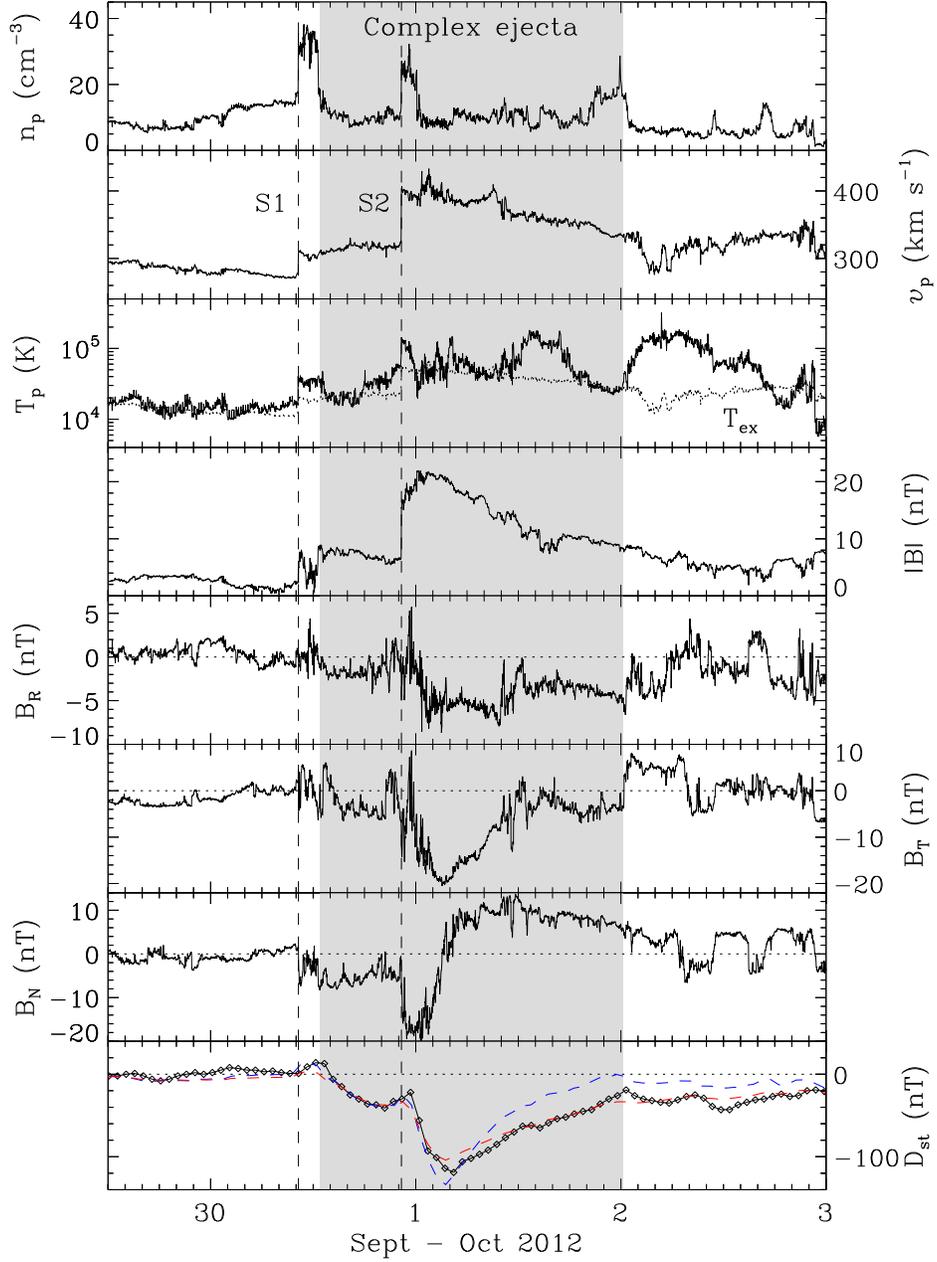} 
\caption{Solar wind measurements at Wind and associated $D_{\rm st}$ index for the 2012 September CME-CME interaction event. Similar to Figure~4. The shaded region indicates the interval of the complex ejecta formed by the merging of the September 25 and 27 CMEs. The vertical dashed lines mark the associated shocks. Note that we have updated the $D_{\rm st}$ estimate of \citet{liu14b} using the southward field components in GSM coordinates.}
\end{figure}

\clearpage

\begin{figure}
\epsscale{0.8} \plotone{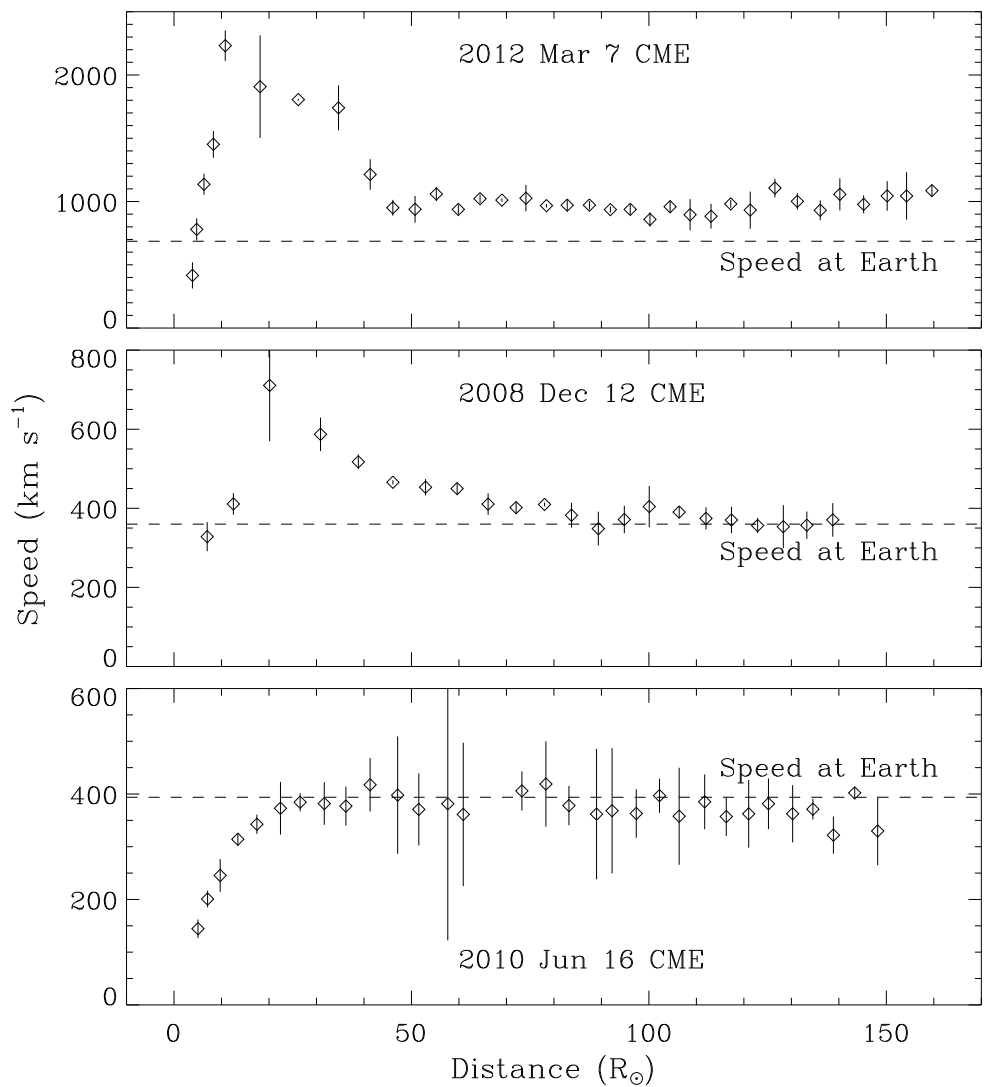} 
\caption{Comparison of Sun-to-Earth propagation profiles between a typical fast CME (upper), a typical intermediate-speed one (middle) and a typical slow one (lower). The horizontal dashed line indicates the observed speed at the Earth. The speed-distance profile of the fast CME, which occurred on 2012 March 7, is adapted from \citet{liu13}. The intermediate-speed CME occurred on 2008 December 12, and its speed-distance profile is after \citet{liu10b}.}
\end{figure}

\end{document}